\documentclass[12pt]{article}
\pdfoutput=1

\usepackage[applemac]{inputenc}

\usepackage{jheppub}
\usepackage{ifpdf}

\usepackage{amssymb}
\usepackage{amsfonts}
\usepackage{color}
\usepackage{graphicx,psfrag, subfigure}
\usepackage{amsmath}
\usepackage{hyperref}
\usepackage[applemac]{inputenc}
\usepackage{dsfont}

\usepackage{soul}
\definecolor{dm}{cmyk}{.20, 0, .30, 0}
\sethlcolor{dm}

\numberwithin{equation}{section}

\newcommand{\vf}{\vec  \phi}
\newcommand{\vk}{\vec k }

\def\be{\begin{equation}}
\def\ee{\end{equation}}
\def\bea{\begin{eqnarray}}
\def\eea{\end{eqnarray}}

\def\M{M_{Pl}}

\def\Ms{M^2_{Pl}}

\def\FRPs{random Fourier potentials}
\def\FRP{random Fourier potential}

\def\Dt{s}

\newcommand{\nn}{\nonumber}
\newcommand{\lp}{\left(}
\newcommand{\rp}{\right)}

\def\M{M_{Pl}}
\def\Ms{M^2_{Pl}}

\newcommand{\ex}[1]{\langle #1 \rangle}

\begin{document}

\begin{titlepage}

\begin{flushright}
SU/ITP-13/13 \\
\end{flushright}

\setcounter{page}{1} \baselineskip=15.5pt \thispagestyle{empty}

\bigskip\
\begin{center}
{\Large \bf  Charting an Inflationary Landscape \\
\vskip 5pt
with Random Matrix Theory}
\vskip 5pt

\vskip 15pt
\end{center}
\vspace{0.5cm}
\begin{center}
{
\large
M.C.~David Marsh,$^1$ Liam McAllister,$^2$ Enrico Pajer$^3$ and Timm Wrase$^4$}
\end{center}

\vspace{0.1cm}

\begin{center}
\vskip 4pt
\textsl{
$^1$ Rudolf Peierls Centre for Theoretical Physics, University of Oxford,\\
1 Keble Road, Oxford OX1 3NP, UK \\
$^2$ Department of Physics, Cornell University, Ithaca, NY 14853, USA \\
$^3$ Department of Physics, Princeton University, Princeton, NJ 08544, USA\\
$^4$ Stanford Institute for Theoretical Physics, Stanford University, Stanford, CA 94305, USA}

\emailAdd{david.marsh1@physics.ox.ac.uk}
\emailAdd{mcallister@cornell.edu}
\emailAdd{epajer@princeton.edu}
\emailAdd{timm.wrase@stanford.edu}

\end{center} %\vfil
{\small  \noindent  \\[0.2cm]
\noindent
We construct a class of random potentials for $N \gg 1$ scalar fields
using non-equilibrium random matrix theory, and then
characterize multifield inflation in this setting.
By stipulating that the Hessian matrices in adjacent coordinate patches are related by Dyson Brownian motion,
we define the potential in the vicinity of a trajectory.
This method remains computationally efficient
at large $N$, permitting us to study much larger systems than has been  possible with other constructions.
We  illustrate the utility of our approach with a numerical study of inflation
in systems with up to $100$ coupled scalar fields.
A significant finding is that eigenvalue repulsion sharply reduces the duration of inflation near a critical point of the potential: even if the curvature of the potential is fine-tuned to be small at the critical point, small cross-couplings in the Hessian  cause the curvature to grow  in the neighborhood of the critical point.}

\vspace{0.3cm}

\vspace{0.6cm}

\vfil
\begin{flushleft}
\small \today
\end{flushleft}
\end{titlepage}

\newpage
\tableofcontents

\section{Introduction}

Cosmic inflation \cite{Guth:1980zm, Linde:1981mu, Albrecht:1982wi} is a successful paradigm for describing the
origin of the primordial density perturbations that are responsible for the large-scale structure of the universe.
The recent results of the Planck satellite \cite{PlanckInflation} are consistent with the broad predictions of single-field inflationary models (for a review, see \cite{Baumann:2009ds}), but there are strong reasons to consider models with multiple scalar fields.

First of all, principles of minimality and simplicity have had mixed success  in the course of discovery of the standard model of particle physics.   In this light it is prudent to  explore a  wide range of possible models for the  inflationary sector.   Second, if one traces the renormalization group flow to higher and higher energies, more and more degrees of freedom appear, and these may well include multiple scalar fields.
Finally, string theory  provides  clear motivation for considering inflationary models  that have {\it{many}}  scalar fields:  presently-understood flux compactifications  based on Calabi-Yau manifolds generally yield  four-dimensional effective theories containing tens or hundreds of  moduli scalars.   The large number is simply a consequence of the rich topology of these spaces.   Inflationary scenarios constructed in string theory can involve a correspondingly large number of  scalar fields that are light enough to fluctuate during inflation.

Significant  efforts have been directed at understanding  the dynamics and signatures of multifield inflation. Although the problem for an arbitrary number of fields was formulated long ago \cite{Starobinsky:1986fxa},  most practical advances  from the past decade concern models with two or three dynamical fields,  or in some cases $N \gg 1$  fields with  a common symmetry structure  (as in e.g.~assisted inflation \cite{Liddle:1998jc}) that greatly simplifies the problem.
The question of the inflationary dynamics  in a field space of dimension $N \gg 1$, with a general potential, has remained intractable.   Analytical tools are scarce, and existing numerical approaches have computational cost that
grows rapidly with $N$.

\subsubsection*{Ensembles of multiple-field  potentials} To assess the typical  features of inflation in general potentials, one may construct large ensembles of potentials and study the dynamics statistically (as done in e.g.~\cite{Tegmark:2004qd} for single-field inflation).
A guiding principle is universality: in high-dimensional  field spaces,  one may expect that large $N$  central limit behavior will lead to results that are independent of the details of the ensemble, and hence to emergent simplicity.
In this paper we will present  a novel and rather powerful technique for  constructing  and studying random potentials for $N \gg 1$ fields.
We will find that the inflationary dynamics in our new ensemble of potentials is largely determined by large $N$ features, which can be expected to extend universally  to a much broader class of potentials than those constructed in this paper.

To obtain an ensemble of low-energy effective potentials, $V(\phi_1,\ldots \phi_N)$, for $N$ real scalar fields, one could in principle begin with an  ensemble of ultraviolet theories in which the field content and symmetries  are specified, but the  interaction strengths are stochastic variables.
In the corresponding  ensemble of low-energy effective theories, the Wilson  coefficients are randomly distributed.
Direct  investigation of such a  Wilsonian landscape  via Taylor expansion is challenging at large $N$: the computational complexity  grows  exponentially with the system size (see discussion below),  so that it is difficult to identify large $N$ asymptotic trends through numerical experiment. Moreover, much of the interesting structure of random functions defined through truncated Taylor expansions tends to arise close to the boundary of the domain of convergence, making  direct examination problematic.\footnote{For example, the zeros (and thereby in some sense the `structure') of the ensemble of single variable degree $k$ complex Gaussian random  polynomials (i.e.~Kac polynomials) concentrate on the unit circle  as $k\rightarrow\infty$ \cite{Kac1, Kac2, Hammersley, SV}.}

Fortunately,  many physically interesting questions  can be addressed without knowing the potential everywhere in  field space.  For example,  inflationary evolution  depends only on the properties of the potential very near to the inflaton trajectory,  and is insensitive to the unexplored regions, so statistical properties of inflation  in an  ensemble of potentials can be deduced by  characterizing the potential along trajectories.

 \begin{figure}[]
   \begin{center}
\label{fig:Intro}
       \includegraphics[width=0.46\textwidth]{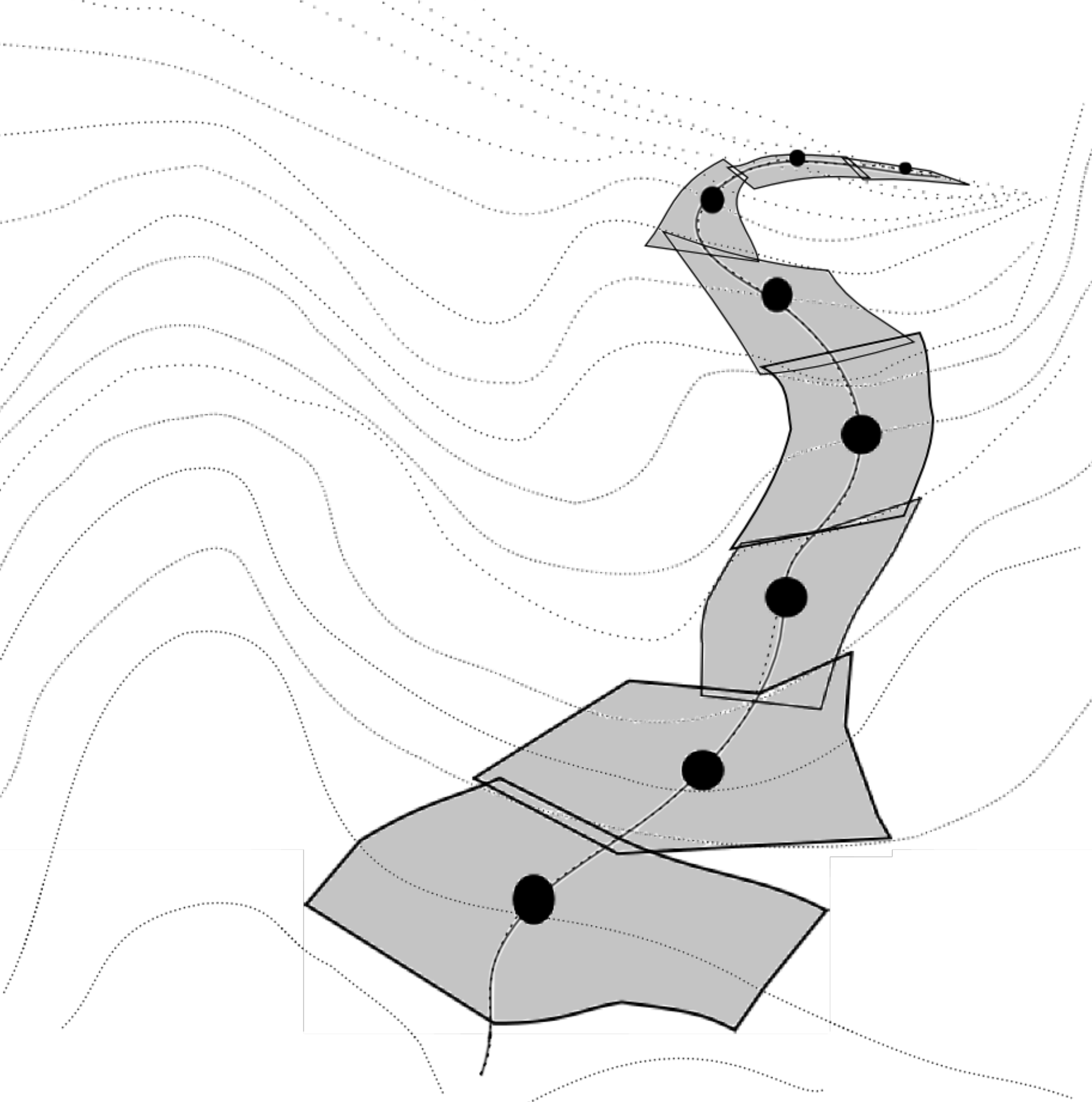}\hspace{.5cm}
    \end{center}
    \caption{By gluing nearby coordinate charts together, a potential along a path $\Gamma$ can be defined.}
\end{figure}

\subsubsection*{A novel, local approach to random potentials} In this paper, we will present a new way of defining random functions \emph{locally} around a path in field space: for a given
path $\Gamma$ in field space,\footnote{We will discuss the restrictions on $\Gamma$ in \S\ref{sec:def}: in particular, it must not self-intersect.} we first specify the values of the potential $V$, gradient $V'$, and Hessian matrix $V'' \equiv {\cal H}$ at a point $p_0 \in \Gamma$. The values of the potential and the gradient vector at a nearby point $p_1 \in \Gamma$  that is separated from $p_0$ by a small path length $\delta s$ may then be obtained to leading order in Taylor expansion from the (known) values of the potential and its first and second derivatives at $p_0$.
The key element of our proposal is to  specify the Hessian matrix at $p_1$ by adding a random matrix to the Hessian at $p_0$,
 \be
 {\cal H}(p_1) = {\cal H}(p_0) + \delta {\cal H} \, , \label{eq:dH}
 \ee
where we have yet to define the  statistical distribution of the  random symmetric matrix $\delta {\cal H}$. By repeating this process along the entire path, we obtain a random function defined in the vicinity of $\Gamma$. In the $\delta s \rightarrow 0$ limit,  one obtains a continuous description of the evolution of the Hessian.

The symmetries of the effective theory impose important constraints on the
distribution of $\delta {\cal H}$.  In this paper we will consider effective potentials $V(\phi_1,\ldots \phi_N)$ in which no direction in field space is special, i.e.~we consider ensembles in which the distributions of the Taylor coefficients
respect statistical isotropy.
We will furthermore assume the existence of a characteristic correlation length, $\Lambda_h$, in field space, and  we will refer to points separated in field space by several units of this `horizontal scale' as \emph{well-separated}.  As for the distribution of the Hessian matrix, our assumptions are:

\begin{enumerate}
\item
The collection of Hessian matrices  associated to a collection of well-separated points along $\Gamma$ constitutes a random sample of a statistical ensemble that is invariant under orthogonal transformations. This is a consequence of statistical rotational invariance.
\item At each point $p \in \Gamma$,
 the $N(N+1)/2$ entries of the Hessian matrix ${\cal H}$  are statistically independent.\footnote{As we explain in \S\ref{sec:comp},  potentials that are constructed via truncated Fourier series do {\it{not}} have this property.}
This  is a  plausible (but not inevitable) property of a landscape of effective theories arising from
 a collection of  sufficiently complicated ultraviolet theories with many degrees of freedom.
\end{enumerate}

These two requirements uniquely restrict the distribution of the Hessian matrix at well-separated points to Wigner's Gaussian Orthogonal Ensemble (GOE) \cite{CambridgeJournals:2039696, mehta2004random}.
The distribution of the eigenvalues  in this ensemble (corresponding to the masses-squared appearing in the Hessian)  is  well-known to be given by a joint probability density function that can be interpreted in terms of a stationary Coulomb gas of $N$ mutually interacting electrically charged particles in two space dimensions, restricted to the real line and confined by a quadratic potential.

As a brief aside, let us mention that while we work exclusively with the GOE in this paper, there are good reasons to expect that many of our results apply for more general ensembles that violate either statistical rotational invariance (and thereby point 1 above), or the assumed independence of the matrix elements (and thereby point 2). Famously, large $N$ universality of random matrix theory  makes the spectral density of matrix ensembles  very robust against the influence of correlations between the entries, as long as the distributions have appropriately bounded moments.
Strong universality results hold even for non-rotationally invariant  distributions, as recently reviewed in \cite{Erdos}.

%%%%%%%%%%%%%%%%%%%%%%%%%%%%%%%%%%%%%%%%%%%%

\subsubsection*{Random potentials generated by Dyson Brownian Motion}
With this motivation, we seek a stochastic evolution of the Hessian matrix along $\Gamma$, as in \eqref{eq:dH}, that gives rise to independent GOE Wigner matrices at well-separated points. Infinitesimally, this evolution will be determined by the law  governing the incremental matrix perturbation $\delta {\cal H}$. This law will necessarily be stochastic, and must furthermore be constructed so that the elements of ${\cal H}$ do not grow indefinitely along $\Gamma$.
A canonical --- albeit to our knowledge not necessarily unique ---  choice for such a process
is obtained  by  stipulating that each matrix element of ${\cal H}$ undergoes an independent Brownian motion between the nearby points
$p_0$ and $p_1$,
\be
\delta {\cal H}_{ab} = \delta A_{ab} + F_{ab}({\cal H}) \frac{\delta s}{\Lambda_h} \, ,
\ee
where $\delta A_{ab}$ are $N(N+1)/2$ zero-mean stochastic variables. Here $F_{ab}({\cal H})$ is a restoring force ensuring that the distribution of the entries of the Hessian remains finite, i.e.\  that the variance does not grow without bound as one travels over  many correlation lengths.\footnote{We note, however, that the boundedness of the elements of the Hessian matrix does not automatically guarantee a similar boundedness for the  magnitudes of the potential  and the gradient vector.}
To obtain the GOE for well-separated points, $F_{ab}$ is uniquely determined to be  \cite{1962JMP.....3.1191D}
\be
F_{ab} = -  {\cal H}_{ab} \, .
\ee
Such a stochastic evolution is called \emph{Dyson Brownian motion} \cite{1962JMP.....3.1191D}, and was
first introduced as an extension of the
stationary Coulomb gas to non-equilibrium configurations.
While Dyson's original formulation introduced a fiducial `time' to parameterize the evolution of the eigenvalues, we here interpret this `time' as the field space path length, $s$,
along $\Gamma$.\footnote{Throughout this paper we will denote this fiducial `time' by $s$ to distinguish it from the proper  Friedmann-Robertson-Walker time, which we will denote by $t$.}

We will use Dyson Brownian motion to specify the evolution of the entries of the Hessian matrix. This fixes the first two moments of $\delta {\cal H}$  to
\bea
\langle \delta {\cal H}_{ab} \rangle &=& - {\cal H}_{ab}(p_0)~\frac{\delta \Dt}{\Lambda_h} \, , \nonumber \\
\langle (\delta {\cal H}_{ab})^2\rangle &=& (1+\delta_{ab})~ \frac{\delta \Dt}{ \Lambda_h} \sigma^2 \label{eq:DBM} \, .
\eea
Here  $\sigma$ denotes  the standard deviation of the corresponding Wigner ensemble.
All higher moments are further suppressed for $\delta s\ll 1$. By repeating this process for a series of nearby points, the potential in the vicinity of $\Gamma$ can be fully specified.

%%%%%%%%%%%%%%%%%%%%%%%%%%%%%%%%%%%%%%%%%%%%

\subsubsection*{Inflation with 100 coupled fields} Thus far, we have kept the path $\Gamma$ arbitrary, but for studies of inflation in random potentials there is a canonical choice: \emph{the inflationary trajectory} in field space.    After specifying the  value of the potential, its gradient, and the Hessian matrix at $p_0$, as well as the field space velocity
$\dot \phi^a$ at the time $t_0$ when the field is at $p_0$, one can solve the coupled Klein-Gordon and Friedmann equations over a short time, $\delta t$. At the time $t_0 + \delta t$, the field will have evolved to some new point $p_1$ close to $p_0$, and we can apply equation \eqref{eq:dH} to find the new Hessian, gradient and value of the potential at $p_1$. By repeating this process, the evolutionary dynamics determines the path $\Gamma$ in field space along which Dyson Brownian motion of the Hessian generates a random potential.

Our construction of a random potential using Dyson Brownian motion is very economical, in the sense that only the bare minimum of information about the potential is retained at each step along the
trajectory. This statement can be made more quantitative by comparing to  alternative constructions of random potentials.
One way to generate  a random potential of a small number of fields  is to  take the potential to be a weighted sum of the first few elements of some basis of the space  of smooth functions,  with the weights being random coefficients.   For example, one could consider a superposition of a number of Fourier modes in a truncated Fourier series, with coefficients drawn from a statistical distribution: see e.g.~\cite{Tegmark:2004qd, Frazer:2011tg,Frazer:2011br, Battefeld:2012qx} for examples of this method.
In this way a random potential is generated globally, though the interesting behavior is generally restricted to a subset of the domain.
The difficulty  in this approach is that a very large number of terms is needed at large $N$.
For a Fourier-expanded potential, the number of terms equals the number of $N$-dimensional wavevectors $\vec k$.  After fixing some UV and IR cutoffs $\vec k_{max}$, $\vec k_{min}$, one  finds that the total number of terms scales as $(k_{max}/k_{min})^{N}$, i.e.~exponentially in the number of fields $N$. On the other hand, the number of terms  needed to study a path in
a Dyson Brownian motion random potential scales only polynomially in $N$: the number of patches required to evolve a potential to a given precision grows linearly with $N$, and upon expanding the potential in any local chart up to second order we have in total only $\mathcal{O}(N^{3})$ terms per inflationary trajectory, which for large $N$ is a huge gain in computational speed.

As a large number of coordinate patches  can be glued together, our construction allows for studies of paths extending  over several correlation lengths, and even several  Planck lengths.\footnote{The implications of random matrix theory for inflation near a critical point, in  systems with $N \gg 1$  fields, were  discussed in \cite{Aazami:2005jf,EM05}.   These  works  made use of a random matrix model for the Hessian evaluated at a  critical point, i.e.~restricted to a {\it single} coordinate patch,  while here we follow a complete trajectory.} In this paper, we illustrate these points by numerically studying  inflation in systems up to size $N=100$, which is an order of magnitude larger than what has been achieved  to date in the context of Fourier-expanded random potentials in cosmology.

%%%%%%%%%%%%%%%%%%%%%%%%%%%%%%%%%%%%%%%%%%%%

\subsubsection*{Inflationary regimes}

We find it useful to distinguish two different parameter regimes:
\begin{itemize}
\item $\Lambda_{h}\ll \M$ : In this regime the potential has features over sub-Planckian distances in field space, so that at a typical point it is too steep to drive inflation. There are two atypical situations in which prolonged inflation can nevertheless take place. The first (see \S\ref{ss:crit} for details), which we dub \textit{inflating near a critical point}, occurs when all the negative eigenvalues of the Hessian happen to fluctuate up to be positive or nearly positive (so that $\eta\ll1$), in the close neighborhood of an approximate critical point (so that $\epsilon\ll1$), with the height of the potential being typical. The second situation (see \S\ref{ss:hs} for details), which we dub \textit{inflating down a high slope}, occurs when both the first and second derivatives take typical values, but the potential takes an atypically large value, so that again both $\epsilon$ and $\eta$ are small. In both these situations, inflation takes place over several patches and ends long before reaching the final minimum.
\item $\Lambda_{h} \gtrsim  \M$ : In this regime,  inflation does not end before reaching the minimum, and the approach to the minimum contributes a sizable fraction of the final 60 e-folds.
We discuss aspects of this regime in \S\ref{sec:LargeField}, but  a complete analysis is difficult with the present implementation of Dyson Brownian motion,  and we leave a  full treatment for the future.
\end{itemize}

This paper is organized as follows: in \S\ref{rmtequilibrium}, we review some key aspects of (stationary) random matrix theory and Dyson Brownian motion.
In \S\ref{sec:Pot} we  present our construction of a class of random potentials,  and  compare our method to the truncated Fourier series approach. In \S\ref{sec:Inflation} we study inflation in this class of potentials, and in \S\ref{sec:concl} we outline some future directions and conclude.

%%%%%%%%%%%%%%%%%%%%%%%%%%%%%%%%%%%%%%%%%%%%%%%%%%%%%%%%%%%%%%%%%%%%%%%%%%%%%%%%%%%%%%%%

\section{Random Matrix Theory in and out of Equilibrium} \label{rmtequilibrium} \label{sec:rmt}
Before presenting our new class of random potentials obtained from non-equilibrium random matrix theory, we will
briefly review
some well-known properties of the stationary  Gaussian Orthogonal Ensemble of real,  symmetric random matrices.

%%%%%%%%%%%%%%%%%%%%%%%%%%%%%%%%%%%%%%%%%%%%%%%%%%%%%%%%%%%%%%%%%%%%%%%%%%%%%%%%%%%%%%%%

\subsection{The Stationary Coulomb Gas}

For the GOE, the joint probability density of  the eigenvalues $x_a$ for matrices of size $N$ is given by
\be
P(x_1, \ldots , x_N) = {\cal C}_{N}~e^{- \frac{1}{2} W} \, , \label{eq:Peq}
\ee
where  the constant ${\cal C}_{N}$ is independent of the eigenvalues, and
\be
W= \frac{1}{\sigma^2} \sum_{a=1}^N x^2_a - \sum_{a\neq b} \ln|x_a - x_b| \, . \label{eq:W}
\ee
In the Coulomb gas formulation of random matrix theory, the eigenvalues $x_a$ are interpreted as the positions of particles confined to the real line, $-\infty < x_a <\infty$, and subject to forces derived from the potential $W$. The  partition function is given by
\be
{\cal Z}_{GOE} = \left(\prod_{a=1}^N \int_{-\infty}^{+\infty} dx_a\right)~ P(x_1, \ldots , x_N) \label{eq:Zeq}\, .
\ee
By expressing equation \eqref{eq:Zeq} in terms of the empirical eigenvalue density,
\be
\rho(\lambda) \equiv \frac{1}{N}\sum_{a=1}^N \delta(\lambda-x_a) \, ,
\ee
one can obtain the famous Wigner semi-circle distribution,
\be
\rho(\lambda) = \frac{1}{\pi N \sigma^2} \sqrt{2 N \sigma^2 -\lambda^2} \, , \label{eq:semicircle}
\ee
from \eqref{eq:Peq} by saddle-point evaluation.  The distribution  \eqref{eq:semicircle} corresponds to the time-independent, macroscopic equilibrium configuration of the thermal system in the  large $N$ limit.

\subsubsection{Fluctuation Probabilities} \label{sec:fluct}
\begin{figure}[ht!]
   \begin{center}
       \includegraphics[width=0.46\textwidth]{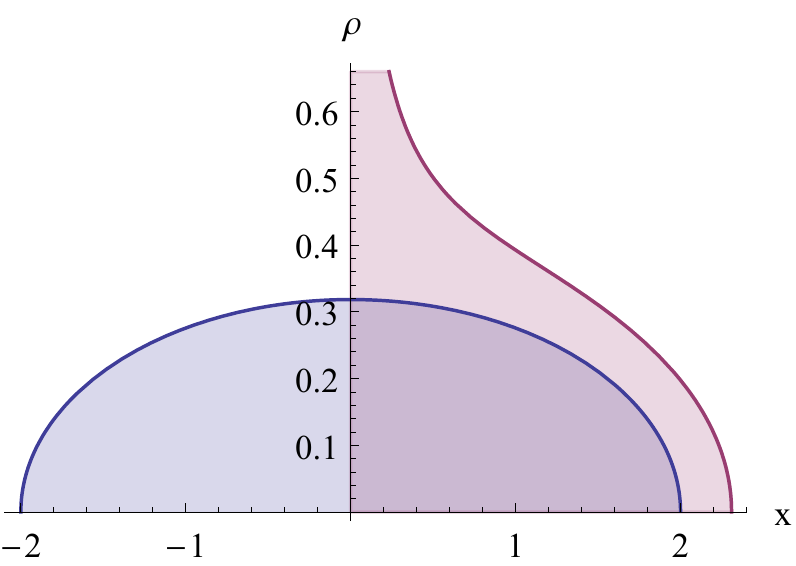}
    \end{center}
    \caption{The spectra for the unfluctuated and fluctuated GOE.}
   \label{fig:WignerSpectra}
\end{figure}

While equation \eqref{eq:semicircle} describes the eigenvalue density of typical configurations of the GOE, interesting sub-samples of the ensemble can be obtained by imposing constraints on the partition function. For instance, the large deviation principle of Ben Arous and Guionnet  \cite{BenArous} states that in the GOE, configurations in which the smallest eigenvalue is greater than or equal to $\xi$ occur with a frequency of
\be
P(x_{min} \geq \xi) = A \exp(-N^2 \Psi_-(\xi)) \, , \label{eq:Pfluct}
\ee
where $\Psi_-(\xi)$ is the `rate function',
\be
\Psi_-(\xi) = {\rm inf}_{\rho_{\xi}} \left(\int_{\xi}^{\infty} \frac12 x^2 \rho_{\xi}(x) dx - \int_{\xi}^{\infty} \int_{\xi}^{\infty} \ln|x-y| \rho_{\xi}(x) \rho_{\xi}(y) dy dx   \right) \, ,
\ee
$A$ is a constant, and we have set $\sigma=\sqrt{2/N}$.
The corresponding fluctuated eigenvalue density $\rho_{\xi}(x)$ can be determined by saddle-point evaluation of the stationary partition function, conditioned on the smallest eigenvalue being no smaller than $\xi$.  One finds \cite{Dean:2006wk, Dean:2008}
\be
\rho_{\xi}(y) =\frac{1}{4\pi}\left( \frac{L(\xi)- y}{y}\right)^{1/2}\left(L(\xi) + 2 y + 2 \xi\right) \, , \label{eq:fluct}
\ee
where $y= x-\xi$, and
\be
L(\xi) = \frac{2}{3} \left[\sqrt{\xi^2 +12} - \xi \right]\, .
\ee
This eigenvalue density corresponds to the
thermal equilibrium configuration of the sub-sample of the GOE that has no eigenvalue smaller than $\xi$. We note in particular the $1/\sqrt{y}$ divergence of the eigenvalue density close to the `edge' at $y=0$. In figure \ref{fig:WignerSpectra} we show the unfluctuated spectrum \eqref{eq:semicircle} for $\sigma^2=2/N$, and the fluctuated spectrum \eqref{eq:fluct} for $\xi=0$.

The corresponding rate function entering equation \eqref{eq:Pfluct} is given by
\be
\Psi_-(\xi) = \frac{1}{432} \Bigl(72\xi^2 - \xi^4 + (30\xi +\xi^3) \sqrt{12 + \xi^2} + 216 \left( \ln 6 - \ln[-\xi + \sqrt{12+\xi^2}]  \right) \Bigr) \,  . \label{eq:Psiminus}
\ee

\subsection{Dyson Brownian Motion} \label{sec:DBM}
We will now briefly summarize the formalism of Dyson Brownian motion, which provides an out-of-equilibrium extension of the Coulomb gas picture of a random matrix ensemble.
It is natural to try to extend the Coulomb gas formulation by promoting the eigenvalues to physical particles with inertia in the Newtonian sense, but Dyson  has argued that this is  difficult or impossible.  On the other hand, the idea of  Dyson Brownian motion \cite{Dyson}, in which
the eigenvalues $x_a$ are interpreted as the positions of particles undergoing \emph{Brownian} motion while subjected  to the deterministic forces derived from the eigenvalue potential \eqref{eq:W},  has proven not only sensible, but also  very useful for obtaining many analytical results in random matrix theory \cite{Erdos}.

In Dyson Brownian motion, the equation of motion of one particle  (i.e., eigenvalue) is given by
\be
\frac{d^2 x_a}{d\Dt^2} = - \frac{dW}{dx_a} -f \frac{dx_a}{d\Dt} + A_a(\Dt) \, , \label{eq:EOM}
\ee
where $A_a(\Dt)$ corresponds to a stochastic, fluctuating force satisfying
\bea
\langle A(s_1) A(s_2) \ldots A(s_{2n+1})\rangle &=& 0 \, , \\
\langle A(s_1) A(s_2) \ldots A(s_{2n})\rangle &=& \sum_{\rm pairs} \langle A(s_a) A(s_b) \rangle \langle A(s_c) A(s_d) \rangle \ldots \, ,
\eea
and
\be
\langle A(s_1) A(s_2) \rangle = \frac{2 }{ f} \delta(s_1-s_2)  \, ,
\ee
where we have suppressed the subscript $a$ for the different independent Brownian motions. The term  $-f \frac{dx_a}{d\Dt}$ represents a frictional force resisting the motion of the particles, and $W$ denotes the eigenvalue potential, cf.~equation \eqref{eq:W}. The `time' coordinate $\Dt$ introduced in \eqref{eq:EOM} corresponds to a fiducial time, which we will later interpret as measuring proper distance  along paths in the scalar field space.

\begin{figure}[ht!]
   \begin{center}
    \subfigure[$N=5$]{\label{fig:EigenvalueRelaxationN=5}
       \includegraphics[width=0.46\textwidth]{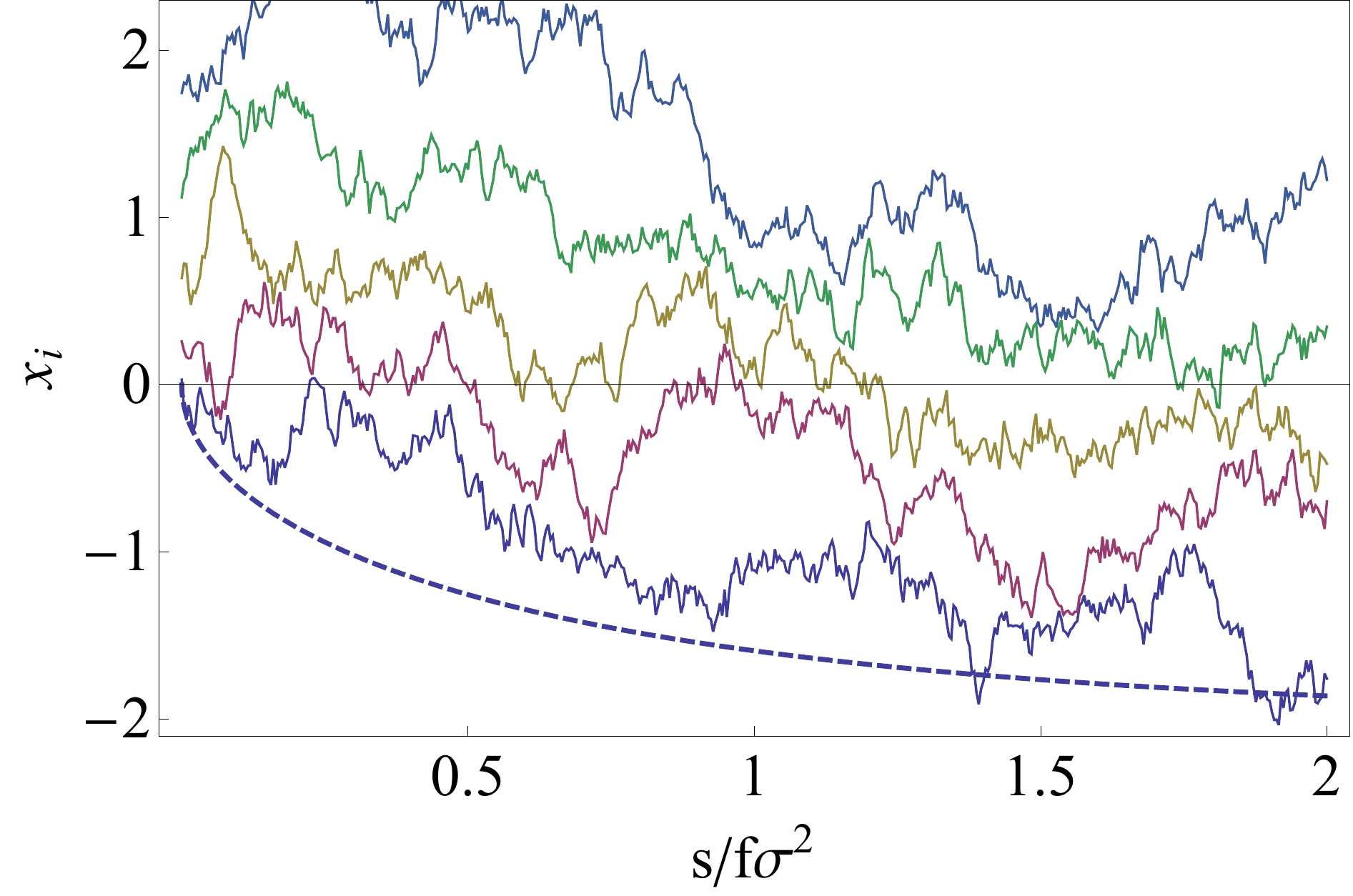}}\hspace{.5cm}
     \subfigure[$N=50$] {\label{fig:EigenvalueRelaxationN=50}
        \includegraphics[width=0.46\textwidth]{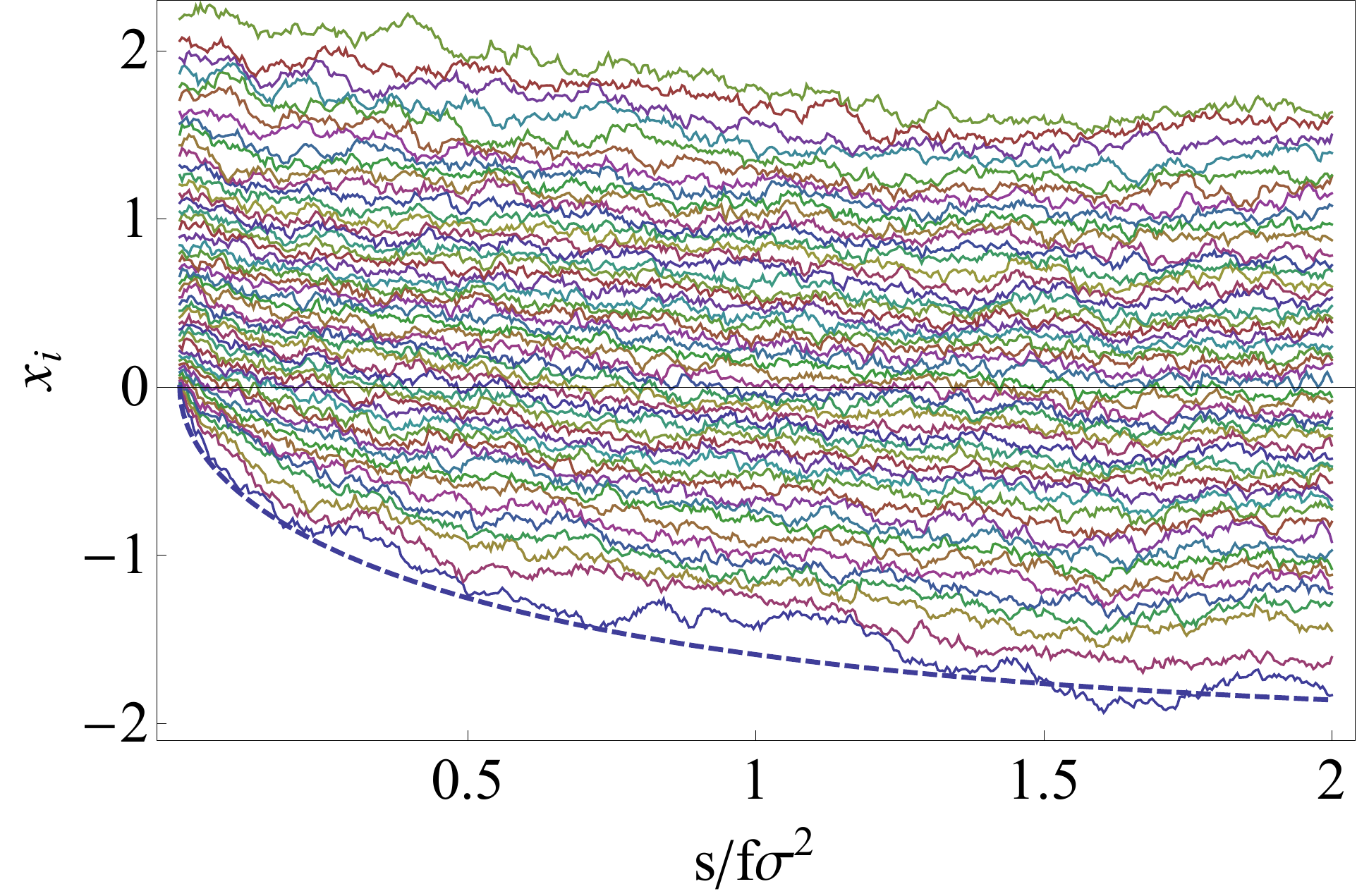}}
    \end{center}
    \caption{Eigenvalue relaxation for an initially fluctuated mass spectrum with smallest eigenvalue $x_{min} = 0$. The dashed line indicates the approximate  mean of the smallest eigenvalue, cf.~equation  \eqref{eq:Bound2}.   The spectrum has been normalized so that the unfluctuated eigenvalues are in the range $[-2,2]$ by setting $\sigma^2 = 2/N$ and  $f=2/\sigma^2$. }
   \label{fig:relax}
\end{figure}

The Brownian motion equation \eqref{eq:EOM} for the eigenvalues of a matrix\footnote{The elements of $M$ can be thought of as dimensionless numbers.} $M_{ab}$ arises from a time-dependent perturbation to the matrix  that relaxes any initial eigenvalue configuration to a configuration consistent with the equilibrium configuration over the timescale $\sigma^2 f$.  Specifically, a matrix $M^0_{ab}$ at time $\Dt=0$ is perturbed by Dyson Brownian motion to the matrix $M_{ab} = M^0_{ab} + \delta M_{ab}$ at time $\delta \Dt$, where the small random perturbation $\delta M_{ab}$ satisfies
\bea
\langle \delta M_{ab} \rangle &=& - M^0_{ab}~\frac{\delta \Dt}{\sigma^2 f} \, , \nonumber \\
\langle (\delta M_{ab})^2 \rangle &=& (1+\delta_{ab})~ \frac{\delta \Dt}{f}  \, .
\eea
All higher-order correlations are further suppressed in $\delta \Dt/\sigma^2f$. The Dyson Brownian motion gives rise to a Smoluchowski/Fokker-Planck equation for the time-dependent probability distribution for the matrix $M(\Dt)$, which for a deterministic initial condition $M(0)=M^0$ is solved by \cite{Uhlenbeck:1930zz}
\be
P(M, s) = {\cal C} \frac{1}{(1-q^2)^{\frac{N(N+1)}{4}}} \exp\left[
-  \frac{{\rm tr} \left(M(\Dt)- qM^0\right)^2}{2\sigma^2(1-q^2)}
\right] \, , \label{eq:Pt}
\ee
with $q=\exp[-\Dt/(\sigma^2 f)]$, which --- as is illustrated in figure \ref{fig:relax} --- quickly  tends to the equilibrium density \eqref{eq:Peq} as $\Dt/(\sigma^2f) \rightarrow \infty$. Thus, the quantity $\sigma^2 f$ serves as a `correlation time' and a measure of the time over which initial conditions become forgotten in the random matrix ensemble. The normalization constant ${\cal C} = (2\pi \sigma^2)^{-N(N+1)/4}$ ensures that
\begin{equation}
\int d^{N(N+1)/2}M~P(M(s)) =1 \, ,
\end{equation}
at all times.

\subsubsection{Eigenvalue Relaxation}\label{sec:EigRelaxation}
For applications to inflation, we will be particularly interested in configurations in which the most tachyonic direction is atypically small in magnitude. Correspondingly, we will be interested in the relaxation over time of the smallest eigenvalue of a fluctuated eigenvalue configuration. We note that the time-dependent partition function corresponding to \eqref{eq:Pt} is at any fixed time $s$ given by,
\be
{\mathcal Z}_{\rm DBM}(s) = \int d M(s)~P(M(s), s) \, , \label{eq:Zofs}
\ee
which only depends on $\Delta M(s) = M(s) - q M^0$. Since the measure transforms trivially under the change of variables from $M(s)$ to $\Delta M(s)$, we find that the partition function \eqref{eq:Zofs} of the time-dependent ensemble coincides with that of the stationary GOE (cf.~equation \eqref{eq:Zeq}) for the matrix $\Delta M(s)$. Thus, the spectrum of $\Delta M(s)$ is given by a Wigner semicircle with a time-dependent, growing, standard deviation,
\be
\sigma^2_{\Delta M} = \sigma^2 \left(1-q^2\right) \, .
\ee

On the other hand, the smallest eigenvalue of the matrix $M(s)$ is bounded by the sum of the smallest eigenvalues of $M^0$ and $\Delta M(s)$ as
\be
\lambda_{min}[M(s)] \geq q\lambda_{min}[M^0]  + \lambda_{min}[\Delta M(s)] \, .\label{eq:bound2}
\ee
While the smallest eigenvalue of $M^0$ is given by a deterministic initial condition, the smallest eigenvalue of $\Delta M(s)$ will be randomly distributed with Tracy-Widom distribution \cite{Tracy:1992rf}. Since the mean of this distribution does not surpass the lower `edge' of the semi-circle law, we find for $\sigma=\sqrt{2/N}$ that
\be
\langle \lambda_{min}[\Delta M(s)]\rangle \geq  -2 \sqrt{1-q^2} \, ,
\ee
and we may bound the \emph{mean} of the smallest eigenvalue at any time $s$,
averaged over all Dyson Brownian motions emerging from $M_0$ at $s=0$, by
\be
\langle \lambda_{min}[M(s)] \rangle \geq q\lambda_{min}[M^0]  -2 \sqrt{1-q^2} \label{eq:Bound} \, .
\ee
However, the bound \eqref{eq:bound2} is almost never saturated, since this would require that the smallest eigenvalues of $M^0$ and $\Delta M(s)$  correspond to exactly the same eigenvector.
Thus, the bound \eqref{eq:Bound} is likewise always respected, but practically never saturated. From numerical simulations one can extract a more accurate, albeit less rigorous, approximation for the evolution of the smallest eigenvalue. For sufficiently large $N$ and for $\lambda_{min}[M^0] \approx 0$ we find that the data is well fitted by the simple expression
\be
\langle \lambda_{min}[M(s)] \rangle \approx -2 \sqrt{1-q} \label{eq:Bound2} \, .
\ee
Equation \eqref{eq:Bound2} will allow us to estimate the rate at which eigenvalue repulsion increases the curvature of a potential that has been fine-tuned to be flat at one point.
In figure \ref{fig:relax} we illustrate how \eqref{eq:Bound2} gives a good indication of the relaxation of the smallest eigenvalue of $M(s)$.
%While equation \eqref{eq:Bound} constitutes a firm bound for the \emph{mean} of the smallest eigenvalue of the matrix $M(s)$, it does not bound fluctuations of the smallest eigenvalue at finite $N$ and, as is illustrated in figure \ref{fig:relax}, somewhat rare fluctuations may cause the smallest eigenvalue to surpass the bound on the mean.

Thus, we see that  Dyson Brownian motion relaxes the eigenvalues of $M(s)$ to configurations consistent with the Wigner semicircle law on the timescale $\sigma^2 f$. Correspondingly, the Brownian motion efficiently mixes the
eigenvectors
of $M(s)$, so as to uncorrelate eigenvectors over the same timescale. Specifically, in the eigenbasis of $M_0$, the normalized eigenvectors $e^{(i)}_a(s)$ of $M(s)$ at times $s\gg \sigma^2 f$ are `delocalized' in the sense that their \emph{l}$^p$ norms,
\be
|| e^{(i)} ||_p = \sum_{a=1}^N (e^{(i)}_a)^p \, ,
\ee
for $p>2$ are significantly smaller than unity. At large times, most eigenvectors become completely delocalized, with  \cite{Erdos2009}
\be
||e^{(i)}||_p \sim N^{\frac{1}{p}-\frac{1}{2}} \, .
\ee

\section{Random Potentials from Dyson Brownian Motion} \label{sec:Pot}
We now show that Dyson Brownian motion can be used to construct a class of random scalar potentials. This construction has very wide applicability that is not restricted to cosmology, but to illustrate the utility of the method we will use these random potentials in \S\ref{sec:Inflation} to study inflation with many fields.  While here we consider potentials realizing statistical rotational invariance, the general method can be extended to other classes of potentials, including the supergravity potentials relevant for approximately-supersymmetric compactifications of string theory \cite{Marsh:2011aa, Chen:2011ac, Bachlechner:2012at}.

\subsection{Constructing Random Potentials via Dyson Brownian Motion} \label{sec:def}

As discussed in the introduction, our method is based on  gluing a long series of nearby coordinate patches together
in order to
follow the potential along a given path over distances much larger than the size of any patch, and in fact larger than the correlation length of the potential. Concretely, we consider a system of $N$ scalar fields $\phi^a$ subject to a potential $V(\phi^{a})$. Around an arbitrary point $p_0$ in field space, the potential can be expanded up to second order as
\bea\label{potV} \label{eq:potV}
V&=& \Lambda_{v}^{4} \sqrt{N} \left[v_0  +  v_a \tilde \phi^a + \frac{1}{2}  v_{ab} \tilde \phi^a \tilde \phi^b  \right]\,,
\eea
where  $\Lambda_{v}$ denotes the `vertical scale', which is determined by  the value of the potential at $p_0$.
Here  $\tilde \phi^a \equiv \phi^a/\Lambda_{h}$ are  dimensionless fields (local coordinates close to $p_0$), with $\Lambda_{h}$ being a `horizontal scale', i.e.~correlation length of the potential.
With these conventions, $v_0$, $v_a$, $v_{ab}$, and $\tilde\phi$ are all dimensionless, and  the overall factor $\sqrt{N}$ has been introduced for reasons we will discuss below.

Let us now discuss how, given the potential in some patch around $p_0$, the potential is constructed in an adjacent patch, i.e.~around a point $p_1$ that lies close to $p_0$ with local coordinates $\delta\tilde \phi^a$ satisfying $ ||\delta\tilde \phi^a|| \ll 1$ in the
Cartesian vector norm.
The potential at $p_1$ may again be written in the form \eqref{eq:potV} in terms of local coordinates around $p_1$,  with
\bea
v_0 \big|_{p_1} &=& v_0 \big|_{p_0} + v_a \big|_{p_0} \delta \tilde \phi^a \, ,  \label{v01}\\
v_a \big|_{p_1} &=& v_a \big|_{p_0} + v_{ab} \big|_{p_0} \delta \tilde \phi^b \, ,  \label{va1}\\
v_{ab} \big|_{p_1} &=& v_{ab} \big|_{p_0} + v_{abc} \big|_{p_0} \delta \tilde \phi^c \, , \label{eq:DiscreteEvol}
\eea
to linear order in $\delta \tilde \phi^a$.
Here the $v_{abc}$  can obviously be interpreted as Taylor coefficients in the expansion of  a globally-specified function $V(\phi^{a})$.   However, we will  soon take a different approach, defining the $v_{abc}$ as appropriate random variables.
The potential can be subsequently updated from one nearby point to the next along any desired path $\Gamma$
in field space: at the $n$th step the potential and its derivatives take the form (\ref{v01})-(\ref{eq:DiscreteEvol}) with $p_0 \to p_{n-1}$ and $p_1 \to p_{n}$.

We now \emph{define} the class of \emph{Dyson Brownian motion random potentials} by stipulating that the perturbations $\delta v_{ab}\big|_{p_n} \equiv v_{abc}\big|_{p_{n-1}}\delta\tilde \phi^{c}$ to the entries of the Hessian matrix are random variables satisfying the conditions \eqref{eq:DBM} of Dyson Brownian motion, i.e.
\bea
\langle \delta v_{ab}\big|_{p_n} \rangle &=& - v_{ab}\big|_{p_{n-1}}~\frac{ ||\delta \phi^a||}{\Lambda_{h}} \, , \nonumber \\
\langle (\delta v_{ab}\big|_{p_n})^2 \rangle &=& (1+\delta_{ab})~ \frac{||\delta \phi^a||}{\Lambda_{h}} \sigma^{2} \label{eq:Postulate} \, ,
\eea
where again $||\delta \phi^a||$ denotes the Cartesian vector norm of the displacement vector  $\delta \phi^a$, and we have chosen a correlation length in field space, $\Lambda_h$, corresponding to the correlation length of  the Brownian motion $\sigma^2 f$,
\be
\Lambda_h \equiv \sigma^2 f \, .
\ee
In this way, the eigenvalues of the Hessian undergo Dyson Brownian motion as the path is traversed. The random perturbation to the Hessian matrix  subsequently affects both the first derivative of the potential and the value of the potential at points further along the path. The motivation for the factor of $\sqrt{N}$ in equation \eqref{eq:potV}  can now be explained: if ${\cal O}(v_{ab})= 1/\sqrt{N}$, so that the magnitude of a typical eigenvalue of $v_{ab}$ is of order one, then  $v_0$ and $v_a$ will both receive contributions of order $1/\sqrt{N}$ over each correlation length. For potentials that are uncorrelated over distances of order $\Lambda_h$ in field space, it is therefore natural to take ${\cal O}(v_0) = {\cal O}(v_a) ={\cal O}(v_{ab})= 1/\sqrt{N}$.

With the  above prescription, the Hessians at points separated in field space by path lengths much longer than $\Lambda_{h}$ are automatically
well-approximated as independent random elements of the Wigner ensemble, cf.~equation \eqref{eq:Pt}, as might be expected in a random potential. By using Dyson Brownian motion to define a random potential along a path in field space, we can follow the evolution of the potential over distances much larger than the correlation length $\Lambda_h$, or even extending over multiple Planck lengths.

We should emphasize that  while the specific choice (\ref{eq:Postulate})  is well-motivated, and leads to potentials whose Hessian matrices obey GOE  statistics  at large separations in field space,  the general idea of taking $\delta v_{ab}\big|_{p_n}$ to be a stochastic variable  whose  statistical properties determine the nature of the global potential  has much wider validity.   One could construct many different classes of potentials by modifying the rule (\ref{eq:Postulate}) governing the evolution of the Hessian.

\begin{figure}[ht!]
    \begin{center}
        \includegraphics[width=0.83\textwidth]{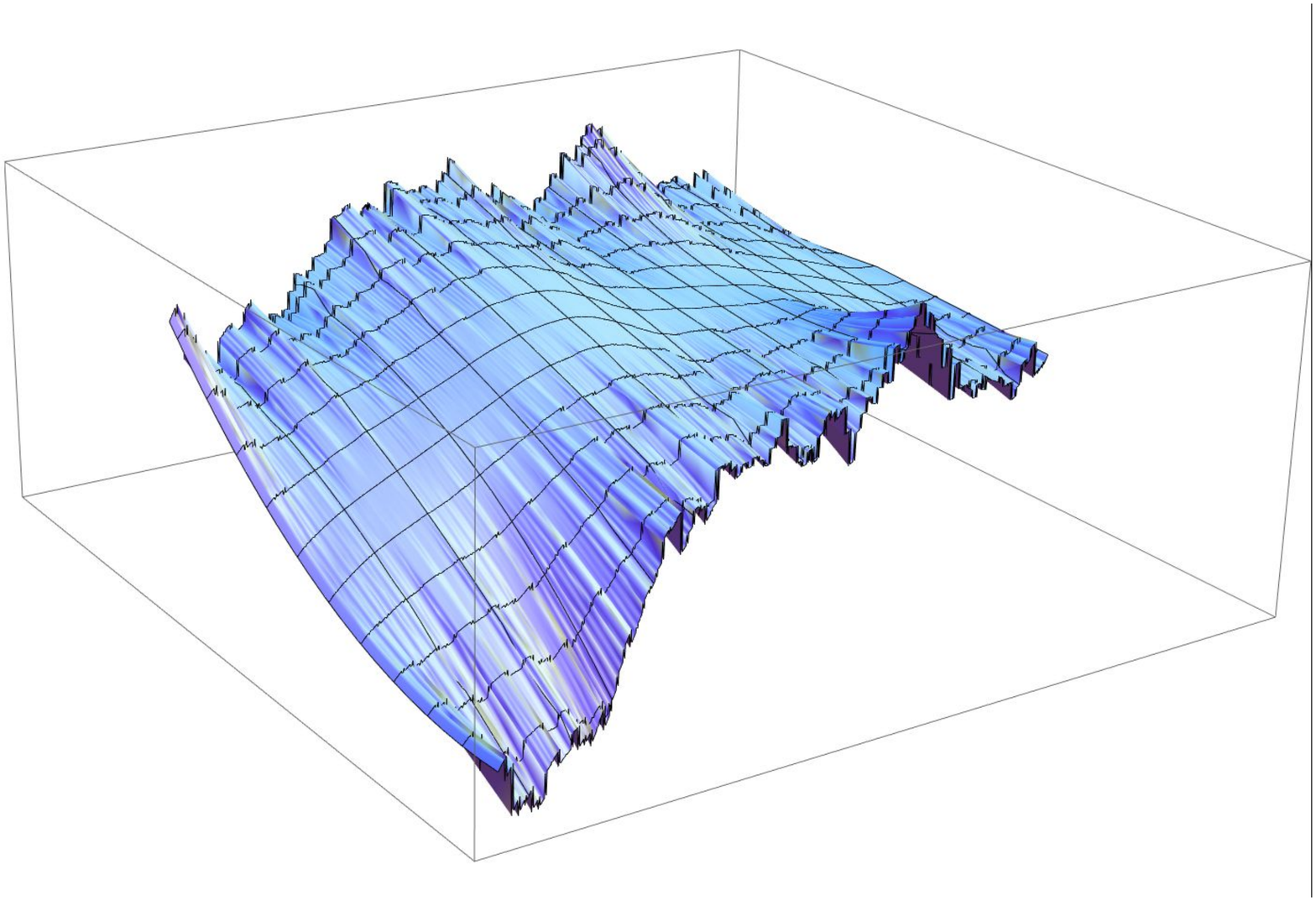} \hspace{.5cm}
    \end{center}
    \caption{The random potentials presented in this paper exhibit non-trivial structure on scales larger than a few $\Lambda_h$, as is illustrated above  for  $N=2$ and a path length of $4 \Lambda_h$.  For illustration purposes we have exaggerated  the separation  between subsequent charts, though a smaller separation will be used in \S\ref{sec:Inflation} to ensure a good approximation to the smooth evolution of the eigenvalues of the Hessian.
    }
   \label{fig:DBMpot}
\end{figure}

Because our method defines the potential in a semi-local as opposed to global fashion, it has some obvious drawbacks. First of all, as the random potential is defined as a sequence of quadratic approximations in a string of coordinate patches, the global structure of the potential far from the generating path is not readily available with this method. Constraints on the structure of the potential from e.g.~Morse theory are therefore not immediately applicable to these potentials.

Furthermore,  the path length along a curve $\Gamma$ does not always give a good measure of distance in field space. In particular, self-intersecting trajectories will  generally not give rise to single-valued potentials. For trajectories  that are nearly self-intersecting --- which is not uncommon for low-dimensional field spaces, but is extremely rare at large $N$ --- a more careful analysis is required.

\subsection{Comparison with Random Fourier Potentials} \label{sec:comp}
It is illuminating to contrast the new type of random potentials we present in this paper to other possible  constructions --- in particular to those determined by  generating random expansion coefficients for a truncated set of basis vectors of the space of smooth functions. The most notable example of this type is that of a Fourier series in which only modes with wave-numbers within a particular interval are included, and these modes have randomly distributed Fourier coefficients. We will refer to potentials of this type as {\it{random Fourier potentials}}. Inflationary cosmology in \FRPs~has been discussed in e.g.~\cite{Tegmark:2004qd, Frazer:2011br, Battefeld:2012qx, Battefeld:2013xwa}, while inflation in multifield ensembles derived in string theory was studied in \cite{Agarwal:2011wm,Dias:2012nf,McAllister:2012am}; for an alternative approach to inflation  in a random potential, see \cite{Tye:2008ef}.

In this section, we will show that our new class of random scalar potentials built on Dyson Brownian motion differs qualitatively from \FRPs.\footnote{More detailed statistical properties of the  class of potentials constructed here will be  presented in a future publication.} While our ensemble of potentials by construction respects the \emph{independence} and \emph{identical distribution} (i.i.d.) of
the
elements of the Hessian matrix, we will see that the Hessian in \FRPs~has  $\mathcal{O}(N^{2})$ correlations between different (diagonal) entries. This general point applies equally to continuous Fourier transforms and discrete Fourier series, which we discuss in turn.

\subsubsection{Construction by Continuous Fourier Transform}

Take $\vf$ to be a vector in $\mathbb{R}^{N}$, where $N$ is the number of fields. Then the potential $V(\vf)$ is a scalar function on $\mathbb{R}^{N}$ that can be Fourier expanded as
\be
V(\vf)=\int \frac{d^{N}k}{(2\pi)^{N}}\, e^{i \vf \cdot \vk} \,V(\vk)\,,
\ee
where the fact that $V(\vf)$ is real imposes the constraint $V^{\ast}(\vk)=V(-\vk)$. We assume that $V(\vk)$ is a Gaussian random field with zero mean and two-point correlation
\be
\ex{V(\vk)V(\vk')}=(2\pi)^{N}\delta^{N} \left(\vk+\vk'\right)\, P(k)\,,
\ee
where $k=|\vk|$ and $P(k)$ is the power spectrum. With this choice, namely the delta function and the fact that $P$ depends only on the norm of $\vk$, the field is statistically homogeneous and isotropic.

We are now interested in the distribution of the matrix elements of the Hessian,
\be
V_{ab}=\frac{\partial^{2} V}{\partial \phi_{a}\partial \phi_{b}}=-\int \frac{d^{N}k}{(2\pi)^{N}}\, k_{a}k_{b}\,e^{i \vf \cdot \vk} \,V(\vk)\,.
\ee
While the one point function of $V_{ab}$ simply vanishes, the two point function satisfies
\be\label{abcd}
\ex{V_{ab}(\vf)V_{cd}(\vf)}=\int \frac{d^{N}k}{(2\pi)^{N}} \, k_{a}k_{b}k_{c}k_{d} \, P(k)\,.
\ee
By rotational invariance, using the fact that the only invariant tensor is $\delta_{ab}$, one finds\footnote{It is easy to check this  by starting with $V(R \vf)$, with $R$ an orthogonal matrix: the Hessian takes the form $R^{T}V_{ab}R$, which  upon substitution in (\ref{lt}) leaves the right hand side invariant.}
\be
\ex{V_{ab}(\vf)V_{cd}(\vf)}=\int \frac{d^{N}k}{(2\pi)^{N}} \, \Bigl( \delta_{ab}\delta_{cd} f_{1}(k)+ \delta_{ac}\delta_{bd} f_{2}(k)+\delta_{ad}\delta_{bc} f_{3}(k) \Bigr) \,\,,\label{lt}
\ee
for some functions $f_{i}(k)$ ($i=1,2,3$) of $k = |\vk|$. The right hand side of \eqref{abcd} is explicitly invariant under permutations of the indices $a,b,c,d$, which implies
\be
f_{1}(k)=f_{2}(k)=f_{3}(k) \, .
\ee
The last two terms of \eqref{lt} (those involving $f_{2,3}$) are indeed present  for an i.i.d. Wigner matrix,  but the first term is a novel feature of \FRPs~that distinguishes their Hessians from those of the GOE. From its index structure, $\delta_{ab}\delta_{cd}$, it follows that the correlation between any two different diagonal elements of the Hessian is non-vanishing and equal to the variance of any other off-diagonal element. We therefore conclude that the Hessian coming from a \FRP~is \textit{not} a Wigner matrix: it features $N(N-1)/2$ correlations among its (diagonal) elements, and therefore violates the assumption of i.i.d.~entries.

Let us discuss  the significance of this discrepancy. On the one hand, strong results on the universality of the spectral density (cf.~\cite{2005math.ph...5003S}), can be used to show that the spectral density of  Hessian matrices in \FRPs~does coincide with the Wigner semi-circle law at large $N$, despite the correlations of equation \eqref{abcd}.
Furthermore, higher-order correlation functions in particular local limits have been shown to exhibit a universal behavior   \cite{Erdos}. On the other hand, general  higher-order correlation functions and large deviation principles (as in equation \eqref{eq:Pfluct}) --- and thereby at least some of the inflationary predictions --- may certainly differ between the two non-equivalent ensembles of Hessian matrices.\footnote{In particular, the correlation among diagonal elements found in \FRPs~but not in GOE landscapes, i.e.~the term involving $f_1(k)$ in \eqref{lt}, could make large fluctuations of the spectrum more probable in \FRPs.}

While we have argued that  a sufficiently complicated, statistically rotational invariant UV theory may plausibly give rise to low-energy effective theories with potentials of the type
described
in this paper, rather than to \FRPs, it is very easy to envision symmetries or correlations in the UV theory that do give rise to correlations in the low-energy theory. In fact, F-term supergravity with a small supersymmetry breaking scale ($m_{3/2}^2 \ll m_{susy}^2$) is of this type \cite{Marsh:2011aa}. While the correlations of the form \eqref{abcd} are not of the form predicted by supergravity, it may well be possible to construct UV theories that give rise to \FRPs~upon RG flow, though we know of no explicit example.

\subsubsection{Construction by Discrete Fourier Series}

For completeness we now describe a \FRP~with an IR cutoff, i.e.\ using a Fourier series instead of a Fourier transform. The potential is taken to be (see e.g. \cite{Tegmark:2004qd,Frazer:2011tg,Frazer:2011br})
\be\label{fs}
V(\vf)=\frac{a_{0}}{2}+\sum_{\vk_{i}}^{\infty} \left( a_{\vk} \cos (\vf \cdot\vk) +b_{\vk} \sin (\vf \cdot\vk) \right)\,,
\ee
where $a_{\vk}$ and $b_{\vk}$ are independent Gaussian random variables with zero mean and some variance $\sigma^{2}_{\vk}$ for each $\vk_{i}$, with $\vk_{i}$ the set of wavevectors.
In the absence of a UV  cutoff,  infinitely many wavevectors are included.

This random potential respects statistical homogeneity, as can be seen by noticing that $V(\vf + {\rm const}) $ can be rewritten in the form \eqref{fs} with
\bea
a_{\vk}\rightarrow \tilde a_{\vk}&=& a_{\vk} \cos(\vk \cdot{\rm const})+b_{\vk} \sin (\vk \cdot{\rm const})\,,\\
a_{\vk}\rightarrow \tilde b_{\vk}&=& b_{\vk} \cos(\vk \cdot{\rm const})-a_{\vk} \sin (\vk \cdot{\rm const})\,,
\eea
where one recognizes that $\tilde a$ and $\tilde b$ are also Gaussian random variables with zero mean and the same variance. Using statistical homogeneity we can restrict ourselves to considering operators at the origin $\vf=0$. For example, the Hessian is
\be
V_{ab}(\vf=0)=-\sum_{\vk_{i}}^{\infty}k_{a}k_{b}a_{\vk}\,.
\ee
The variance is then
\be
\ex{V_{ab}(0)V_{cd}(0)}=\sum_{\vk_{i}}^{\infty} k_{a}k_{b}k_{c}k_{d} \sigma^{2}_{\vk}\,,\label{abo}
\ee
where $\sigma^{2}_{\vk}$ is the variance of $a_{\vk}$, and we have used the delta function coming from the correlator to get rid of the other sum. A rotation $R$ in field space acting as $\vf\rightarrow R \vf$ changes the right hand side to the same expression with $\vk\rightarrow R^{T}\vk$.
For \eqref{abo} to be invariant, it is not sufficient for $\sigma$ to be a function of $|\vk|$, because the discrete sum over $\vk_{i}$ is not invariant under rotations. This can be fixed by going to the continuum limit, or equivalently by taking the periodicity to be larger than any other length in the problem (i.e.\ sending the IR cutoff to infinity).
Rotational invariance can also be violated by the UV cutoff: for example, the choice made in \cite{Frazer:2011tg} of summing $\vk_{i}$ inside a hypercube ($k_{max,a}< \Lambda$ for each $a=1,\dots,N$) rather than a hypersphere ($\sum_{a} k_{max,a}^{2}<\Lambda$) explicitly violates rotational invariance.

\section{Multifield Inflation in Dyson Brownian Motion Potentials} \label{sec:Inflation}
In this section, we  use the Dyson Brownian motion method for defining random potentials to study multifield inflation.

\subsection{Background Evolution} \label{sec:bkg}
As explained in \S\ref{sec:def}, we will consider potentials defined by a quadratic approximation in a series of coordinate patches.   Let us now see how  inflation can occur in potentials of this form.
We will study a system of $N$ real scalar fields $\phi^a$ with canonical kinetic terms (i.e.~the field space is Euclidean $\mathbb{R}^N$) and with equations of motion given by
\bea
\ddot{\phi}^a + 3 H \dot{\phi}^a + \frac{\partial V}{\partial \phi^a} &=& 0\,,  \\
3 H^2 \Ms &=& \frac{1}{2} \sum_{a=1}^N ( \dot  \phi^a )^2 + V(\phi^a) \, .
\eea
To study the field evolution close to an arbitrarily chosen origin in field space, we may truncate the potential at quadratic order, leading to \eqref{potV}, which we repeat here for convenience:
\bea
V&=& \Lambda_{v}^{4} \sqrt{N} \left[v_0  +  v_a \tilde \phi^a + \frac{1}{2}  v_{ab} \tilde \phi^a \tilde \phi^b  \right]\,,\label{eq:Potential}
\eea
It is  helpful to rewrite the equations of motion in terms of dimensionless quantities. For this purpose we will use the number of e-folds $N$ to measure time, via the exact relation $Hdt=dN$. Denoting derivatives with respect to $N$ by primes we have
\bea
\tilde\phi^{a ''}+ (3-\epsilon_H) \tilde\phi^{a'}+\frac{V_{,\tilde a}}{\Lambda_{h}^{2} H^{2}}&=&0\,, \nonumber \\
\frac{V}{M_{pl}^{2}H^{2}}+\frac{1}{2} \left(\frac{\Lambda_{h}}{M_{Pl}}\right)^{2} \sum_{a=1}^N(\tilde \phi^{a'})^{2}&=&3 \label{eq:eom} \,,
\eea
where $\epsilon_H$ denotes the first (Hubble) slow-roll parameter,
\be
\epsilon_H= -\frac{\dot{H}}{H^2} \, ,
\ee
and $V_{\tilde a}=\partial V/\partial \tilde \phi^{a}$.
The overall dimensionful scale of the potential, $\Lambda_v$,  is fixed by  requiring  that the potential satisfies the COBE normalization
\be
\left(\frac{V|_\star}{\epsilon|_\star}\right)^{1/4} = 0.027 \M \, ,
\ee
at the time $t_\star$ at around 60 e-folds before the end of inflation.  Since we will not discuss perturbations in the fields $\phi^a$, the actual value of $\Lambda_v$ does not affect the field evolution and will therefore not be relevant for the rest of this section.

We implement Dyson Brownian motion multifield inflation as follows: starting from the origin in field space with velocity $\tilde \phi^{a'}|_0$ at $t =0$, we evolve the system while letting the Hessian matrix execute Dyson Brownian motion. Numerically, this process may be discretized  into iterations conditioned on  $|| \tilde \phi^a( t) || \leq \delta$ for  $0<\delta\ll 1$, and $V(t) >0 $. In other words, after the field has evolved to a point with coordinates $\delta\tilde \phi^{a}$ located a distance $\delta = ||\delta\tilde \phi^{a}||$ away from the origin in field space, the Hessian is perturbed by Dyson Brownian motion, which in turn induces a shift in the lower order Taylor coefficients as well.  After re-centering the field to the origin in field space, the values of the Taylor coefficients of the potential are given to linear order as in equation \eqref{eq:DiscreteEvol}, which we repeat here:
\begin{eqnarray}
v_{0}\big|_{p_n}&=&v_{0}\big|_{p_{n-1}}+v_{a}\big|_{p_{n-1}} \delta\tilde \phi^{a}\,, \nn\\
v_{a}\big|_{p_n}&=&v_{a}\big|_{p_{n-1}}+v_{ab}\big|_{p_{n-1}}\delta\tilde \phi^{b}\,, \\
v_{ab}\big|_{p_n}&=&v_{ab}\big|_{p_{n-1}}+ \delta v_{ab}\big|_{p_{n-1}} \, ,\nn
\end{eqnarray}
with $\delta v_{ab}$ given as in equation \eqref{eq:Postulate}.
This procedure can be repeated to create a string of coordinate patches along the trajectory of the inflaton.

Ideally, one would study trajectories that terminate in a metastable vacuum with $V\approx 0$. Such paths are rare and --- at distances of a few $\Lambda_h$ away from the vacuum --- differ in no obvious way from other Dyson Brownian motion potentials. Thus, we will find it useful to distinguish inflationary histories based on whether or not the last 60 e-folds occur as the field makes its final approach to the vacuum.
In the former case, which is common for $\Lambda_h \gtrsim \M$ and which we study in \S\ref{sec:LargeField}, the nature of the minimum plays a crucial role.
In contrast, for $\Lambda_h \lesssim \M$ we will see in \S\ref{sec:SmallField} that the field traverses multiple patches during inflation, and that inflation ends
well before the field reaches the final vacuum. In this case, we expect that the shape of the potential during the inflationary epoch is largely independent of the detailed structure of the distant vacuum.

\subsection{Numerical Simulations} \label{sec:results}
We are now ready to discuss the results of numerical simulations of inflation in Dyson Brownian motion potentials. While these simulations are extensive, they are at the same time limited in scope: in this paper we will only discuss   the  \emph{background evolution} of the coupled multiple-field system \eqref{eq:eom} subject to slow-roll initial conditions. Quantum fluctuations of fields in Dyson Brownian motion potentials will be discussed in a subsequent publication.

For the field to start out in slow-roll, we require that $\ddot{\tilde{\phi}}^a(t=0)=0$, and that
\be
\dot{\tilde{\phi}}^a(t=0) = - \frac{1}{3\Lambda_h^2 H} \frac{\partial V}{\partial \tilde{\phi}^a} \ll \frac{\Lambda_v^2 N^{\frac14} \sqrt{2v_0}}{\Lambda_h} \, .
\ee
To characterize the shape of the potential at the initial point, it is useful to introduce the (potential) slow-roll parameters at $\tilde \phi^a=0$,
\be
\epsilon_V (\tilde \phi^a=0) \equiv \frac{\M^2}{2} \frac{\sum_{a=1}^N (\partial_a V)^2}{V^2} = \frac{1}{2}\left( \frac{\M}{\Lambda_h}\right)^2 \frac{\sum_{a=1}^N v_a^2}{v_0^2} \label{eq:eps} \, ,
\ee
and
\be
\eta_V(\tilde \phi^a=0) \equiv \M^2 \frac{ {\rm min(}{\rm Eig}( V_{ab}))}{V} = \left( \frac{\M}{\Lambda_h}\right)^2 \frac{{\rm min(Eig(}v_{ab}))}{v_0} \label{eq:eta} \, .
\ee
We also define \cite{GrootNibbelink:2001qt}
\bea
\eta_{\perp}&=& \frac{\sqrt{(\sum_a V_{,a}^2)(\sum_b (\dot{\phi^b})^2)-\M^5 \sum_a V_{,a} \dot{\phi}^a}}{H\sum_a(\dot{\phi}^a)^2} \nn\\
&=& 3 \lp\frac{\M}{\Lambda_h}\rp^2 \frac{\sqrt{\lp\sum_a v_a^2\big|_{p_n}\rp \lp\sum_b (\tilde{\phi}^{b'}\rp^2 - \lp \sum_a v_a\big|_{p_n}\tilde{\phi}^{a'}\rp^2}}{v_0\big|_{p_n}\tilde{\phi}_{b}^{'} \tilde{\phi}^{b'}}\, , \label{eq:etaperp}
\eea
where in the second line we have used, as mentioned above, that we re-center the field at the origin after each step. For multifield inflation $\eta_{\perp}$ is a very useful quantity, since it characterizes the acceleration of the field in the hyperplane perpendicular to the instantaneous field space velocity.\footnote{For an alternate  parameterization of the slow-roll parameters in multifield inflation, see \cite{Achucarro:2010da}.}

Finally, before discussing  our statistical results, we note that  the form of the potential \eqref{eq:Potential} and the slow-roll parameters \eqref{eq:eps}-\eqref{eq:eta} suggests a convenient separation of these studies into separate regimes. For a given collection of independent random points, it is natural to expect that the values of  the dimensionless quantities $v_0$, $v_a$, $v_{ab}$ are typically of the same order of magnitude.\footnote{The distributions of these quantities will be discussed in detail in a subsequent publication.} For the eigenvalues of $v_{ab}$ to be of order unity, this corresponds to ${\cal O}(\sqrt{\langle v_0^2 \rangle})={\cal O}(\sqrt{\langle v^2_a\rangle})={\cal O}(\sqrt{\langle v^2_{ab} \rangle}) = 1/\sqrt{N}$. Sustained slow-roll inflation may occur when the slow-roll parameters \eqref{eq:eps}-\eqref{eq:eta} are small, which may happen in different ways. First, for $\Lambda_h \gg \M$, the potential is slowly varying over super-Planckian distances, and large field inflation is possible without fine-tuning of $v_a$ and $v_{ab}$. We will discuss this case in \S\ref{sec:LargeField}. Second, for $\Lambda_h \lesssim \M$, small slow-roll parameters require either atypically small values for both $v_a$ and the smallest eigenvalue of $v_{ab}$, or atypically large values of $v_0$.\footnote{Of course,
a combination of these possibilities, involving atypical values for all of $v_0$, $v_a$  and $v_{ab}$ may also support inflation.}  We will now discuss
these cases in turn.

\subsection{Inflation from Many Patches: $\Lambda_h \ll \M$} \label{sec:SmallField}

As mentioned above, for $\Lambda_h\ll \M$, inflation can take place in two different regimes. The first regime concerns regions close to approximate critical points where the gradient and the smallest eigenvalue of the Hessian are small. We will call this regime \textit{inflating near a critical point}. The second regime concerns regions where the potential takes atypically large values. We will call this regime \textit{inflating down a high slope}.  However, before considering inflation in these regimes,  we will estimate which of these possibilities is more likely to occur in some collection of random points.  To make this heuristic estimate, we will assume that at this collection of points,
$v_0$, $v_a$ and $v_{ab}$ are independent, random, Gaussian variables with vanishing mean and with standard deviation $\sigma = 1/\sqrt{N}$. It is worth commenting on this assumption (which we use exclusively in this heuristic estimate). In the present construction, the Dyson Brownian motion potential is not bounded from above or below, and in fact one could reach arbitrarily high or low values of $V$ by moving far enough in one fixed direction.
For the inflationary dynamics considered in this work, implementing upper and lower bounds is unnecessary, and would not change our conclusions, so we leave it for future work.
For the current heuristic estimate we have in mind the fact that non-renormalizable couplings between the vacuum energy and the scalar fields will generically induce masses of order
$ m^2 \sim \ex{V}/\Lambda_h^2 $, where $\ex{V} = v_0 \Lambda_{v}^{4} \sqrt{N}$. This in turn implies that for typical values $v_{ab}\sim v_{0}$, justifying the distributions of $v_{0}$, $v_a$, and $v_{ab}$ assumed above.

The typical values of the slow-roll parameters are therefore
\bea
&\epsilon_V& \sim N  \left(\frac{\M}{\Lambda_h} \right)^2  \, , \\
&\eta_V& \sim \sqrt{N} \left(\frac{\M}{\Lambda_h} \right)^2 \, .
\eea
Both $\epsilon_V$ and $\eta_V$ may be suppressed at points at which $v_0$ takes on atypically large values: for typical values of $v_a$ and $v_{ab}$ but with $v_0 \sim \frac{1}{\bar \eta}  \left(\frac{\M}{\Lambda_h} \right)^2$ the slow-roll parameters are of the order
\bea
&\epsilon_V& \sim \bar \eta^2  \left(\frac{\Lambda_h}{\M} \right)^2  \, , \\
&\eta_V& \sim \bar \eta \, .
\eea
The fraction of random points of this type, with large $v_0$ but typical gradient and curvature, is therefore
\be
\mathcal{F}_1 \sim e^{- v_0^2/\sigma^2} = \exp\left( - \frac{N}{\bar \eta^2}  \left(\frac{\M}{\Lambda_h} \right)^4\right) \, .
\ee
This fraction should be compared with the fraction of random points at which $v_0$ is typical, but the slow-roll parameters are suppressed because the gradient and curvature of the potential are small.
For $v_{a} \sim \frac{\sqrt{\bar \epsilon}}{N} \left(\frac{\Lambda_h}{\M} \right)$ and ${\rm min(Eig}(v_{ab})) \sim \frac{\bar \eta}{\sqrt{N}} \left(\frac{\Lambda_h}{\M} \right)^2$, the slow-roll parameters are
\bea
&\epsilon_V& \sim \bar \epsilon   \, , \\
&\eta_V& \sim \bar \eta \, .
\eea
The likelihood that a randomly chosen point is of this type is given by
\be
\mathcal{F}_2\sim P\left(|v_a| < \frac{\sqrt{\bar \epsilon}}{N} \left(\frac{\Lambda_h}{\M} \right) ~~ \forall a\right)\times  P\left({\rm min(Eig}(v_{ab}) \geq  \frac{-|\bar \eta|}{\sqrt{N}} \left(\frac{\Lambda_h}{\M} \right)^2 \right) \, ,
\ee
where the probability distribution for the gradient is (by assumption) Gaussian,\footnote{To be precise, the probability  that the gradient has small magnitude  is  governed by a chi-squared distribution,  but for simplicity of presentation we use the  estimate given above, which yields the same asymptotic expression: this factor is in any event subleading in importance compared to the probability that the Hessian  eigenvalues fluctuate.} and the fluctuation probability of the smallest eigenvalue of $v_{ab}$ is given by \eqref{eq:Pfluct}. Thus, the fraction of points of this type is of the order
\be
\mathcal{F}_2\sim \left(  \frac{\sqrt{\bar \epsilon}}{N} \left(\frac{\Lambda_h}{\M} \right)\right)^N \times \exp\left[-N^2 \Psi\left(\frac{- |\bar \eta|}{\sqrt{N}} \left(\frac{\Lambda_h}{\M} \right)^2\right)\right] \, .
\ee
As the first factor is subdominant in $N$ for large systems, the probability scales as
\bea
\mathcal{F}_2 &\sim& \exp\left[-N^2 \Psi\left(\frac{- |\bar \eta|}{\sqrt{N}} \left(\frac{\Lambda_h}{\M} \right)^2\right)\right]  \approx  \exp\left[-N^2 \left( \Psi(0) - \frac{|\bar \eta|}{\sqrt{N}} \left(\frac{\Lambda_h}{\M}  \right)^2 \Psi'(0) \right) \right] \nonumber \\
&\sim& \exp\left[-N^2\frac{\ln(3)}{4}\right] \, .
\eea
for $N\gg 1$. Here, we have used that the first Taylor coefficient of the rate function is $\Psi(0) = \ln(3)/4$, cf.~equation \eqref{eq:Psiminus}.

Thus, for $\Lambda_h \leq \M$, inflation is more likely to be driven by a fine-tuned gradient and Hessian matrix, rather than by a large $v_0$, for systems of size
\be\label{eq:Nrange}
1\ll N <  \frac{4}{\ln(3) \bar \eta^2}  \left(\frac{\M}{\Lambda_h} \right)^4\, .
\ee
We restrict our numerical studies to $N \leq 100$, so that \eqref{eq:Nrange} is always satisfied when $\Lambda_h \leq \M$ and $\bar{\eta} \ll 1$. However, since \eqref{eq:Nrange} depends on the particular distributions for $v_0$, $v_a$ and $v_{ab}$ we now study the characteristics of inflation for both cases.

%%%%%%%%%%%%%%%%%%%%%%%%%%%%%%%%%%%%%%%%%%%%%%%%%%

\subsubsection{Inflating Near a Critical Point}\label{ss:crit}
\label{sec:cpinflation}
In models of single field slow-roll inflation, one often finds $N_e \sim 1/|\eta_V|$,  so that fine-tuning of the mass-squared of the inflaton close to an approximate critical point can lead to a significant number of e-folds.
We will now discuss the many-field generalization of this result, and we will show that the  amount
of slow-roll inflation generated close to an approximate critical point is approximately \emph{independent} of the value of $\eta_V$ at the critical point.  This result follows directly from the existence of eigenvalue repulsion, and can therefore be expected to be a generic feature of much broader classes of inflationary potentials than those studied in this paper.

To obtain small slow-roll parameters, we consider inflation close to an approximate critical point ($||v_a||\ll1$) that is also an approximate inflection point (such that ${\rm |min(Eig(}v_{ab}))|\ll1$). More concretely, we consider large sets of points where $v_0 < \lp\frac{\M}{\Lambda_h}\right)^2$,  $v_a$ is small in magnitude and  random in orientation, and $v_{ab}$ is a random member of the subset of the GOE in which no eigenvalue is smaller than $\xi$. Thus, the initial spectrum of eigenvalues at the approximate critical point will be similar to the fluctuated spectrum of figure \ref{fig:WignerSpectra}, except that the left `hard' edge will be at $x=\xi$ rather than at $x=0$.
At this starting point of the field evolution, we consider a diagonalized Hessian matrix and an initial field space velocity $\dot{\phi}^a$ satisfying the slow-roll condition.

As the fields move away from the initial location, the Hessian matrix evolves through Dyson Brownian motion as discussed in \S\ref{sec:Pot}. A distance of $\Lambda_h$ away from the initial point, the eigenvalue spectrum will be essentially uncorrelated with the initial spectrum (cf.~\eqref{eq:Pt}). However, long before the field reaches a distance $\Lambda_h$ from the starting point, any initial fine-tuning of the smallest eigenvalue is ruined by the eigenvalue relaxation process discussed in \S\ref{sec:EigRelaxation}. Using the evolution of the left edge of the eigenvalue density as an indication of the evolution of the smallest eigenvalue (cf.~the discussion around equation \eqref{eq:Bound2}), we find that even when tuning the smallest eigenvalue to zero at the approximate critical point, the rate at which this eigenvalue can be expected to relax to an unfluctuated configuration is given by,
\be
-2\frac{d\sqrt{1-q}}{ds} = -\frac{q}{\Lambda_h \sqrt{1-q}} \, ,
\ee
which diverges as $s \rightarrow 0$. Here again, $q= \exp[-s/\Lambda_h]$.

We believe that this result is of significant conceptual importance for multiple field inflation:  at large $N$, small off-diagonal terms in the Hessian matrix will effectively spoil any initial fine-tuning of the eigenvalues of the Hessian matrix long before one correlation length of the potential is traversed. At the level of the eigenvalues, this is manifest as the  eigenvalue relaxation phenomenon discussed in \S\ref{sec:EigRelaxation}. As a direct consequence, the number of e-folds of multiple field inflation obtained close to an approximate critical point will be approximately independent of the initial slow-roll parameter $\eta_V$, as any tuning of the curvature is promptly erased by the eigenvalue relaxation.\footnote{The number of e-folds will typically still depend on $\epsilon_V$ evaluated at the critical point.}  Moreover, as inflation ends when the slow-roll parameters become of order one, we note that inflation in this regime can be expected to be of small-field type, with most e-folds being generated close to the approximate critical point.

\begin{figure}[ht!]
    \begin{center}
    \subfigure[]{
        \includegraphics[width=0.45\textwidth]{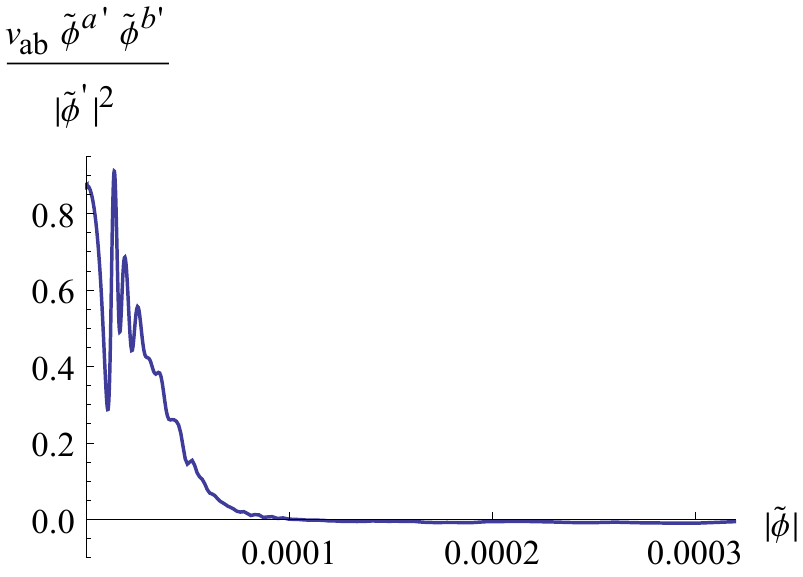}}
     \qquad
     \subfigure[]{
        \includegraphics[width=0.45\textwidth]{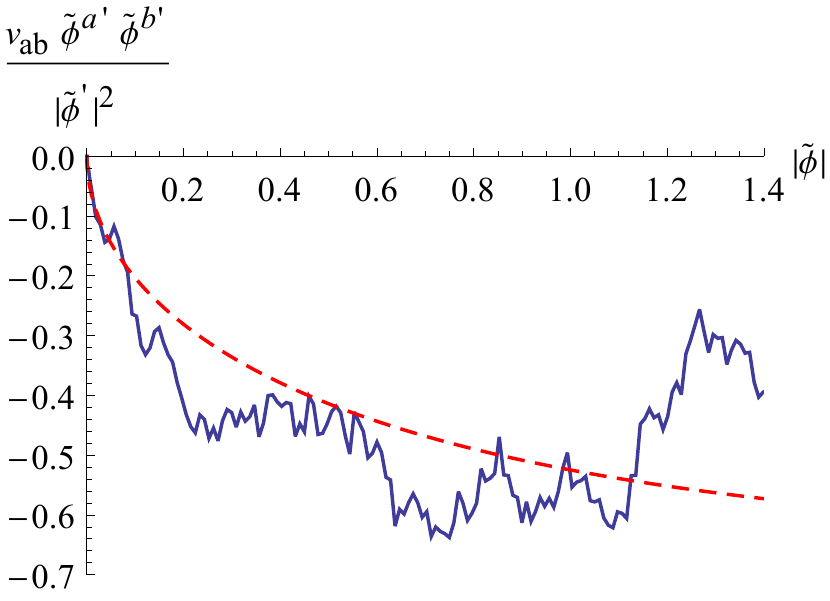}}\\
     \subfigure[]{
        \includegraphics[width=0.45\textwidth]{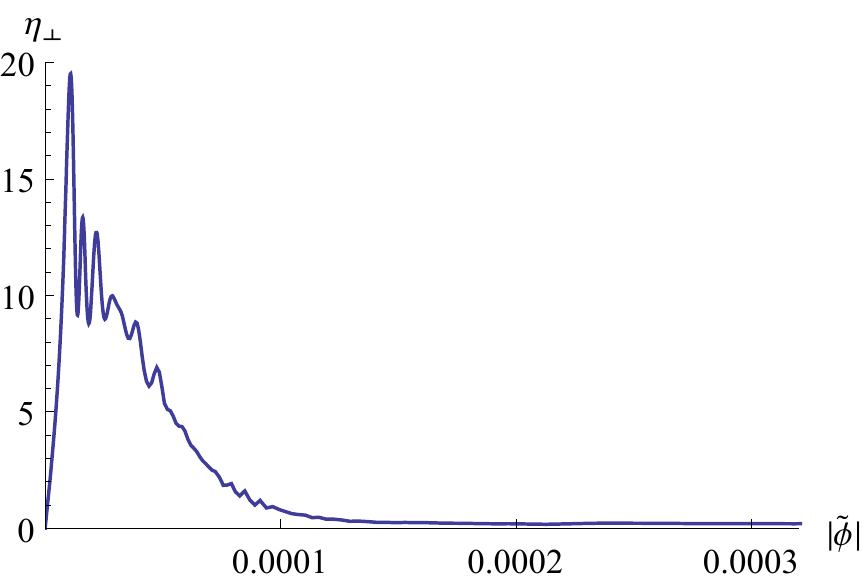}}
     \qquad
     \subfigure[]{
        \includegraphics[width=0.45\textwidth]{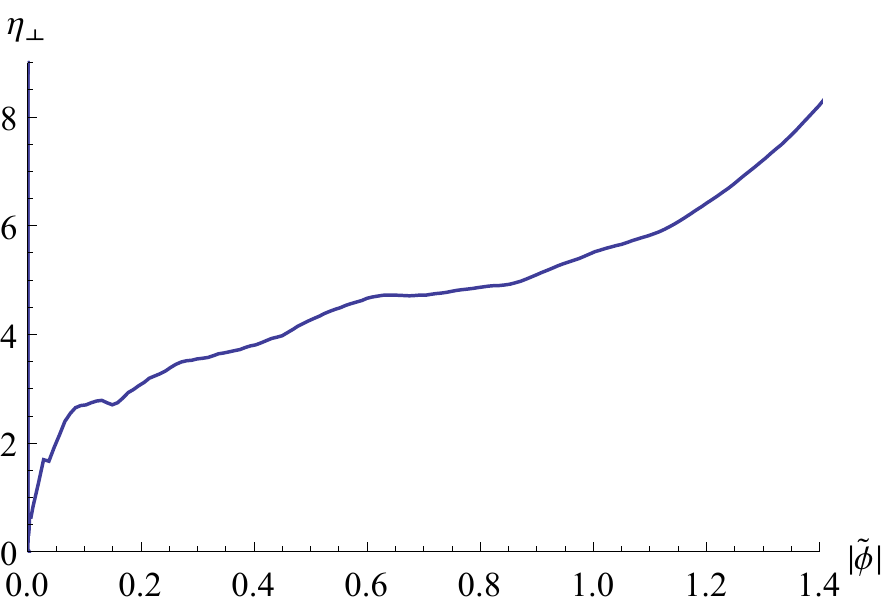}}\
    \end{center}
    \caption{The curvature of the potential, in an example with $N=50$.  Panels (a), (b) show the second derivative of the potential along the instantaneous velocity, while panels (c), (d) show $\eta_\perp$. The second derivative is initially positive, but the inflaton quickly turns towards the (in this case) unique tachyonic direction. The red dashed line in figure \ref{fig:EtaPerpSmallv0}(b) is the analytic model \eqref{eq:VmodelSmallv0}.
    $\eta_\perp$ is initially large while the field turns towards the direction of steepest descent, and then becomes small. Towards the end of inflation all the slow-roll parameters, including $\eta_\perp$, become large.
    The data shown are for an example with $\Lambda_h/\M=.1$, $v_0=10/\sqrt{N}$, $\epsilon_V^{(0)} = 10^{-10}$, and $\eta_V^{(0)} = -10^{-6}$.}
   \label{fig:EtaPerpSmallv0}
\end{figure}

Furthermore, for general initial conditions in which the gradient of the potential at the initial point is not proportional to the eigenvector of the smallest eigenvalue, the inflationary dynamics involves a curved trajectory in field space, and is therefore genuinely multifield. The reason for this behavior is simple: during slow-roll inflation the velocity is antiparallel to the  gradient of the scalar potential, but the gradient vector is not correlated with the eigenvectors of the Hessian, and will therefore receive contributions from multiple eigenvectors as the field moves from the critical point (cf.~equation \eqref{va1} and figure \ref{fig:EtaPerpSmallv0}(a)). Furthermore, as discussed in \S\ref{sec:EigRelaxation}, the eigenvectors of the Hessian matrix delocalize as the field traverses a distance of $\Lambda_h$, thus making initial fine tuning of the direction of the gradient vector inefficient for suppressing turns of the inflationary potential. These effects lead to an initially large value of the parameter $\eta_{\perp}$ introduced in equation \eqref{eq:etaperp}. After this initial turning period, the value of $\eta_{\perp}$ decreases, only to become large again towards the end of inflation, as illustrated in figure \ref{fig:EtaPerpSmallv0}(c), (d).

As we have shown in this section, intrinsically multiple field effects are important for the inflationary dynamics close to an approximate critical point. One may ask, however, whether some of these effects could be captured in a single field toy model.
For obvious reasons, the curving and winding of the multiple field trajectory is hard to capture in a single field model. On the other hand, from the average relaxation of the smallest eigenvalue in equation \eqref{eq:Bound2}, we may construct a model that captures some of the features of the multiple field system.

Denoting the inflaton field in the toy model by $\varphi$ (such that $\tilde \varphi = \varphi/\Lambda_h$), the single-field potential corresponding to an $N$-field model with initial potential determined by $v^{(0)}_0$, $v^{(0)}_a$ and $v^{(0)}_{ab}$, where we introduced the short-hand notation $X^{(0)} = X|_{p_0}$,  can be written as
\be
V = \Lambda_v^4 \sqrt{N} \lp v^{(0)}_0 - v_1 \tilde{\varphi} + \frac{1}{2} v_2 \tilde \varphi^2 \rp
\,,\label{eq:VmodelSmallv0}
\ee
where $v_1 >0$ is related to $v^{(0)}_a$ of the corresponding multiple field model, but is negligibly small since initially $\epsilon_V \ll 1$.

\begin{figure}[ht!]
    \begin{center}
     \subfigure[]{
        \includegraphics[width=0.45\textwidth]{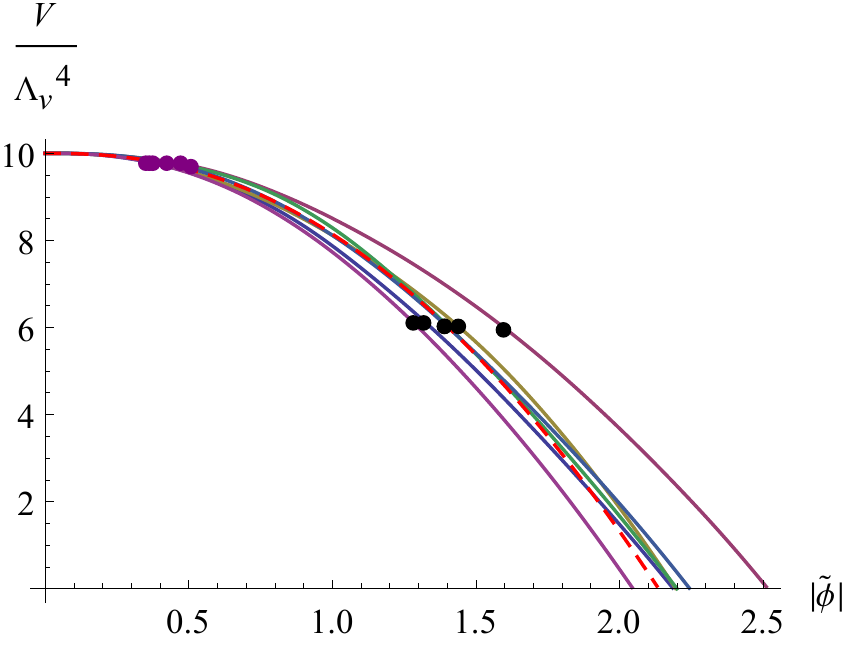}\label{fig:PotentialSmallv0}}
     \qquad
     \subfigure[]{
        \includegraphics[width=0.45\textwidth]{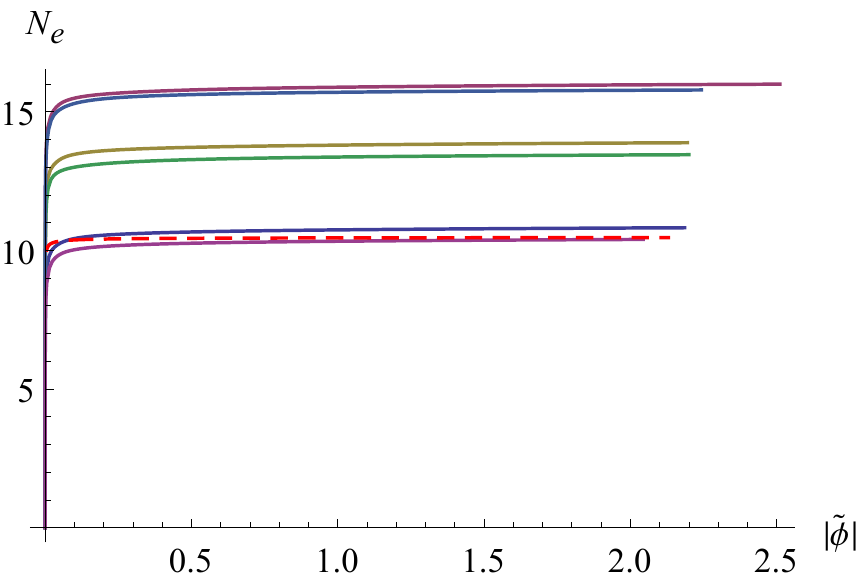}}
    \end{center}
    \caption{The potential and the number of e-folds for six examples with $\Lambda_h/\M=.1$, $v_0=10/\sqrt{N}$, $N=50$, $\epsilon_V^{(0)}=10^{-10}$, and $\eta_V^{(0)} = -10^{-6}$. The dashed red lines correspond to the analytic model \eqref{eq:VmodelSmallv0}. The purple (left) points in figure \ref{fig:PotentialSmallv0} indicate when $\eta_V = 1$ and the black (right) points show when $\epsilon_H=1$.}
   \label{fig:PotNeSmallv0}
\end{figure}

In the full multiple field model, the field quickly starts turning towards a direction along which the second derivative is negative. Thus we expect that the dimensionless mass parameter $v_2$ should be related to the smallest eigenvalue of the Hessian matrix, and thereby to \eqref{eq:Bound2}. Thus, we take $v_2$ to be field-dependent and given by
\be
v_{2}(\tilde \varphi)  \equiv -2 c  \sqrt{1-e^{-|\tilde{\varphi}|}}\,, \label{eigenvaluerelaxationsqrt}
\ee
where we have introduced the parameter $c$ to compensate for the fact that the multiple field configuration does not evolve only in the direction of the smallest eigenvalue. Due to the strong effect of eigenvalue relaxation, we have also neglected any small initial value of $v_2$.
A plot of six realizations of inflation in this regime, with $N=50$, is shown in figure \ref{fig:PotNeSmallv0}, together with the corresponding single field toy-model.

As expected, in this regime most e-folds are generated very close to the approximate critical point, and once the field has evolved a distance $\Lambda_h$ from the starting point, $\eta_V \gtrsim 1$ and inflation is typically no longer supported. Since inflation in this regime occurs at $\varphi \ll \Lambda_h$,
the single-field model can be further simplified to give
\be
V \approx \Lambda_v^4 \sqrt{N} \lp v_0^{(0)} - v_1 \tilde{\varphi}- c |\tilde{\varphi}|^{\frac52}\rp\,, \quad \text{for}\quad |\tilde{\varphi}| \lesssim 1\,.
\ee
The number of e-folds of inflation generated in this toy model is then
\bea
N_e &=&\frac{1}{\M} \int_0^{\varphi_{end}} \frac{d \varphi}{\sqrt{2\epsilon_H}} \approx \frac{\Lambda_h}{\M} \int_0^{\tilde{\varphi}_{end}} \frac{d \tilde{\varphi}}{\sqrt{2\epsilon_V}} = \lp \frac{\Lambda_h}{\M}\rp^2 \int_0^{\tilde{\varphi}_{end}} \frac{|V|}{|V'|} d \tilde{\varphi}\nn\\
&\approx& \lp \frac{\Lambda_h}{\M}\rp^2 \int_0^{\tilde{\varphi}_{end}} \frac{v_0^{(0)}}{v_1  + \frac52 c |\tilde{\varphi}|^{\frac32}} d \tilde{\varphi} \approx \lp \frac{\Lambda_h}{\M}\rp^2 \lp\frac{2}{5 c} \rp^{\frac23} \frac{4 \pi v_0^{(0)}}{3 \sqrt{3} |v_1|^{1/3}}\,,\qquad
\eea
where we used in the second line that $V$ is approximately constant during inflation, and we denoted derivatives with respect to $\tilde \varphi$ by primes.  We have also used that initially $\epsilon_V \ll 1$ and that inflation ends roughly when $\epsilon_V=1$, so that $c |\tilde{\varphi}_{end}|^{\frac32} \gg |v_1|$. Note that (in this limit) the final result is independent of the particular value of $\tilde{\varphi}_{end}$.

This single field model may be further related to the full multiple field system by setting  $|v_1| \approx \tfrac{1}{40} ||v_a^{(0)}||$, where the prefactor is chosen to match the numerical results. That such a prefactor arises can be understood as follows: we initially move in a random direction along which $v_a \tilde{\phi}^a$ is negative, but the mass spectrum has fluctuated so that $v_{ab}$ has hardly any negative eigenvalues.
Thus, initially, before the inflaton starts turning and $v_{ab}$ develops multiple tachyonic directions, $v_a \tilde{\phi}^a$ decreases due to the evolution equation \eqref{eq:DiscreteEvol}.

The number of e-folds can be written in terms of the initial value of the potential $V(0)$ and the initial slow-roll parameter $\epsilon_V^{(0)}$ as
\be
N_e \approx 8 \lp \frac{\Lambda_h}{\M}\rp^{\frac53} \lp \frac{V(0)}{\Lambda_v^4}\rp^{\frac23} N^{-\frac13} \lp \epsilon_V^{(0)} \rp^{-\frac16} \left(\frac{1/3}{c} \right)^{2/3}\label{eq:NeSmallv0}
\,,
\ee
where the value $c\sim1/3$ is motivated by comparison to simulations of the multiple field system. We note that the above formula is independent of $\eta_V^{(0)}$ and is only weakly dependent on $\epsilon_V^{(0)}$.

We find that \eqref{eq:NeSmallv0} agrees very well with simulations of the full multiple field system. We have gathered data for several hundred different values of the parameters in the ranges
\bea
10^{-3} &\leq \frac{\Lambda_h}{\M} &\leq 1\,,\nn\\
10^{-14} &\leq \epsilon_V^{(0)} &\leq 10^{-2}\,,\\
10^{-14} &\leq -\eta_V^{(0)} &\leq 10^{-2}\,,\nn\\
10^{-1} &\leq \sqrt{N} v_0 &\leq 10^{5}\,,\nn
\eea
for $N$ from 10 up to 100. Since we start inflation near a point with small $\epsilon_V^{(0)}$ and $\eta_V^{(0)}$ we find in some rare cases that when we move to a neighboring patch the Dyson Brownian motion leads to a minimum in which  the field becomes stuck, or approximately stuck,  resulting in outliers with a very large number of e-folds that we discard. This is extremely rare except for low $N$ (mostly $N=10$ and $N=20$) and very small $\epsilon_V^{(0)}$ and $\eta_V^{(0)}$. In our numerical simulations we also demand that at each point the inflaton trajectory satisfies $\big|\big|\tilde{\phi}|_{p_n} -\tilde{\phi}|_{p_i}\big|\big| < \big|\big|\tilde{\phi}|_{p_{n+1}} -\tilde{\phi}|_{p_i}\big|\big|$, $\forall i<n$. This means that the distance to any previous patch is monotonically increasing and ensures that our local potential accurately describes a global potential without branch cuts. While this excessive bending occurs more frequently than getting stuck in a minimum, it is  nevertheless rare at large $N$.

\begin{figure}[ht!]
    \begin{center}
         \includegraphics[width=0.5\textwidth]{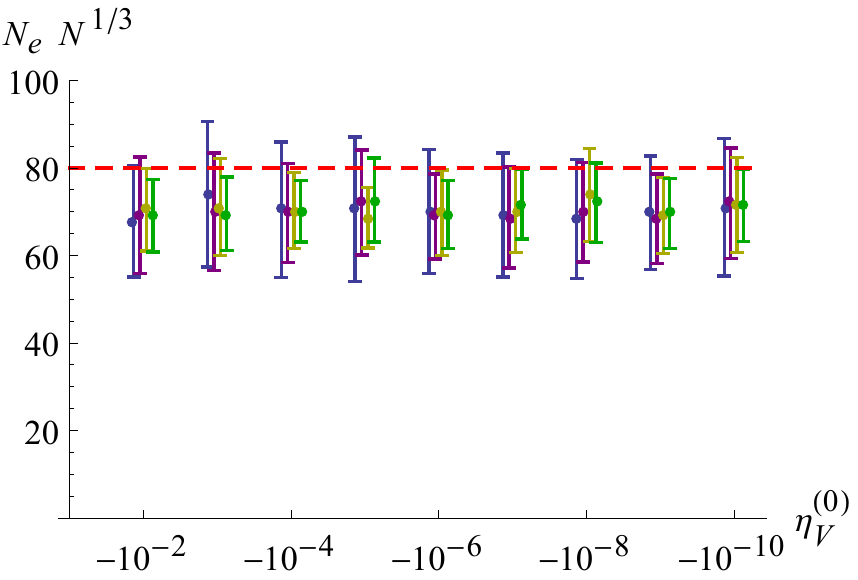}
    \end{center}
    \caption{The number of e-folds $N_e$ multiplied by $N^{1/3}$, together with $1 \sigma$ error bars, as a function of $\eta_V^{(0)}$, for $N=40, 60, 80, 100$ (from left (blue) to right (green)) and for $\epsilon_V^{(0)} =10^{-8}$, $\frac{\Lambda_h}{M_p} = \frac{1}{10}$, $V(0) = 100 \,\Lambda_v^4$. The plot is based on 100 simulations for each $\eta_V^{(0)}$ and $N$ value, and we have slightly displaced the different $N$ along the $\eta_V^{(0)}$ axis for better visibility. The red dashed line is the prediction from our analytic model \eqref{eq:NeSmallv0}.}
   \label{fig:etaindepence}
\end{figure}

In its range of applicability, i.e. for $v_0 < \lp\frac{\M}{\Lambda_h}\right)^2$ and $N_e >1$, we find that \eqref{eq:NeSmallv0} predicts the average number of e-folds fairly well, differing from the numerical value by less than a factor of two. We have also verified that the numerical data scales as predicted by \eqref{eq:NeSmallv0}. For example, in figure \ref{fig:etaindepence} we plot the number of e-folds vs. $\eta_V^{(0)}$ for a variety of $N$ values, showing good agreement with our analytic formula.

%%%%%%%%%%%%%%%%%%%%%%%%%%%%%%%%%%%%%%%%%%%%%%%%%%%

\subsubsection{Inflating Down a High Slope}\label{ss:hs}

As mentioned in the introduction to this section, small slow-roll parameters and a prolonged period of inflation can be obtained for typical values of $v_a$, $v_{ab}$ and very large values for $v_0$. For $v_0 \gg (\M /\Lambda_h)^2$, this gives rise to inflation with an unfluctuated mass spectrum: the spectral density follows the Wigner semi-circle law with  ${\rm min(Eig(}v_{ab}))\approx-2$. In contrast to inflation close to an approximate critical point, Dyson Brownian motion of the Hessian matrix does not immediately spoil any initial fine-tuning, and inflation is sustained  over more than a correlation length, $\Delta\varphi \gg \Lambda_h$.
In this regime, inflation eventually ends when $v_0$ has decreased enough so as to no longer provide sufficient suppression of  the slow-roll parameters. For $\eta_V$, this occurs when $v_0 \lesssim  (\M /\Lambda_h)^2$.

The inflationary dynamics in this regime is again of truly multiple field type, with turning and bending inflationary trajectories. If the initial gradient vector is uncorrelated with the eigenvectors of the Hessian, then the typical inflationary trajectory will turn so as to become aligned with a direction along which the first and second derivatives of the potential are negative.  As approximately half of the eigenvectors of the initial Hessian correspond to tachyonic eigenvalues,  the initial turns in this regime are typically less severe than those observed in the regime of approximately critical point inflation. Furthermore, as $(\M /\Lambda_h)^2 \ll v_0$ we normally have $\eta_\perp < 1$ until the end of inflation, cf.~equation \eqref{eq:etaperp}.

\begin{figure}[ht!]
    \begin{center}
     \subfigure[]{
        \includegraphics[width=0.45\textwidth]{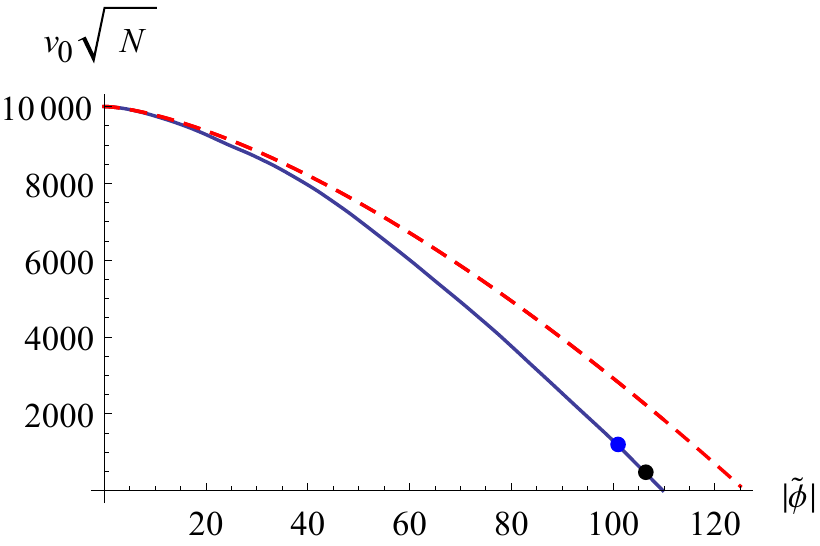}}
     \qquad
     \subfigure[]{
        \includegraphics[width=0.45\textwidth]{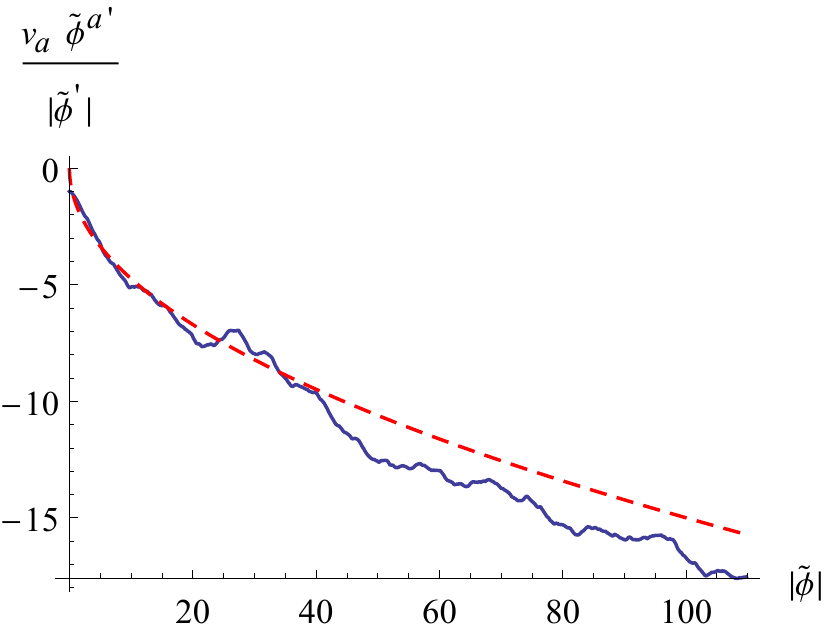}}\\
     \subfigure[]{
        \includegraphics[width=0.45\textwidth]{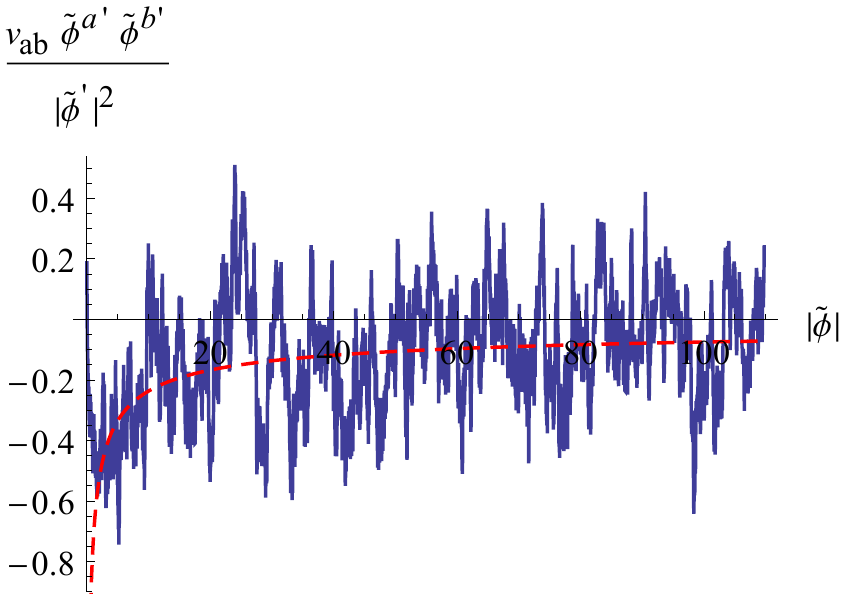}}
     \qquad
     \subfigure[]{
        \includegraphics[width=0.45\textwidth]{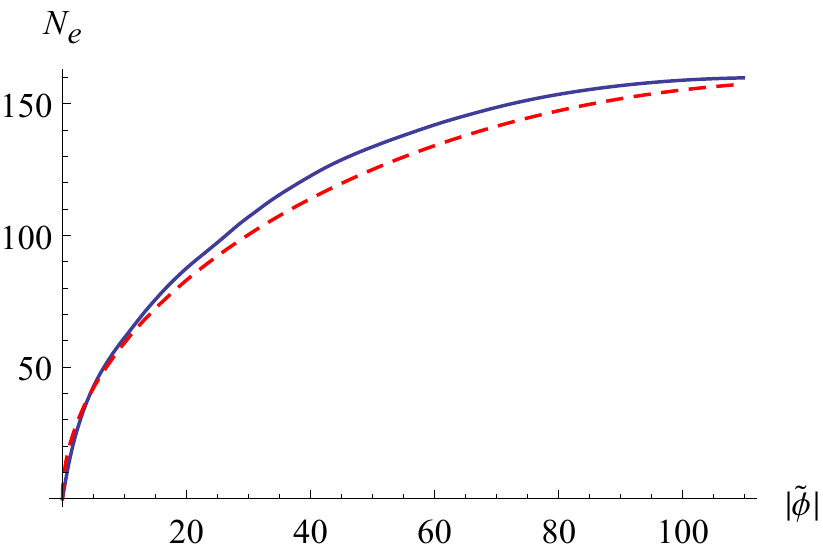}}
    \end{center}
    \caption{Inflation from rolling down a high slope.  These plots show for one example with $\Lambda_h/\M=.1$, $v_0=10,000/\sqrt{N}$, $N=50$ and $\epsilon_V^{(0)} = 2.5 \times10^{-5}$ and $\eta_V^{(0)}=10^{-6}$: (a) the potential along the inflationary trajectory, (b) the first derivative and (c) the second derivative of the potential projected onto the instantaneous velocity of the inflaton, and (d) the number of e-folds. The first (blue) dot in (a) denotes the point where $\eta_V=1$ and the second (black) point denotes the end of inflation where $\epsilon_H =1$. Roughly in between these two points, $\epsilon_V$ and $\eta_H$ become larger than unity. The red dashed lines correspond to our analytic model \eqref{eq:toy2}.}
   \label{fig:largev0}
\end{figure}

As inflation in this regime is sustained over multiple correlation lengths, the eigenvectors of the Hessian can be expected to efficiently mix through eigenvector delocalization, as discussed in \S\ref{sec:EigRelaxation}. It follows that  the projection of the Hessian along the inflaton trajectory will not be constant throughout inflation, and may even become positive.  In these cases,  the component of the gradient vector along the inflationary trajectory will temporarily increase, only to subsequently decrease as the inflaton  adjusts its path towards steeper directions. In general, the projection of the gradient vector onto the inflationary path becomes on average more and more negative throughout inflation, as is illustrated for a particular example in figure \ref{fig:largev0}.

We would again like to obtain a single-field toy model of inflation in this regime, at least for the purpose of estimating the number of e-folds.
From figure \ref{fig:largev0} we see that the instantaneous mass along the field trajectory exhibits complicated oscillations,  which suggests that  the single-field toy model is not well-described by a quadratic potential.  We find that a candidate two-parameter single-field model is given by
\be
V= \Lambda_v^4 \sqrt{N} (v_0^{(0)} - c |\tilde{\varphi}|^p)\,, \label{eq:toy2}
\ee
with $1< p <2$. From comparison to numerical simulation of the multi-field system, $c \sim 1$. In this model --- just as in the full multiple field system --- the magnitude of the gradient grows along the inflationary trajectory.  In \eqref{eq:toy2},  the contribution linear in $\tilde \varphi$ has been consistently neglected, since the potential is dominated by $v_0$ for small $|\tilde \varphi|$, and by $c |\tilde{\varphi}|^p$ for large field values.  In this model, the end of inflation occurs approximately when $\epsilon_V=1$, which happens to leading order in $(\M/\Lambda_h)^2/v_0$ at $V \approx 0$, so that $|\tilde{\varphi}_{end}| \approx (v_0/c)^{\tfrac1p}$.

Calculating the number of e-folds we find
\bea
N_e &=&\frac{1}{\M} \int_0^{\varphi_{end}} \frac{d \varphi}{\sqrt{2\epsilon_H}} \approx \frac{\Lambda_h}{\M} \int_0^{\tilde{\varphi}_{end}} \frac{d \tilde{\varphi}}{\sqrt{2\epsilon_V}} = \lp \frac{\Lambda_h}{\M}\rp^2 \int_0^{\tilde{\varphi}_{end}} \frac{|V|}{|V'|} d \tilde{\varphi}\nn\\
&=& \lp \frac{\Lambda_h}{\M}\rp^2 \int_0^{\tilde{\varphi}_{end}} \frac{v_0^{(0)} - c |\tilde{\varphi}|^p}{c\,p |\tilde{\varphi}|^{p-1}} d \tilde{\varphi}= \lp \frac{\Lambda_h}{\M}\rp^2 \frac{\lp v_0^{(0)} \rp^{\frac{2}{p}}}{2 (2-p)} \lp\frac{1}{c} \rp^{\frac{2}{p}}\,.
\eea
From numerical simulations we find that the shape of the potential and the number of e-folds are captured by taking $c=1$ and $p=\frac32$, in which case
\be\label{eq:NeLargev0}
N_e \approx \lp \frac{\Lambda_h}{\M}\rp^2 \lp v_0^{(0)} \rp^{\frac{4}{3}}\,.
\ee
We have gathered roughly 100 data points for different $N$, $\epsilon_V^{(0)}$ and $\eta_V^{(0)}$ for $\Lambda_h/\M \in \{0.1,1\}$. We find that the average number of e-folds differs from \eqref{eq:NeLargev0} by less than a factor of two as long as $0.1 \leq ||v_a^{(0)} || \leq 5$ and $v_0^{(0)} > (\M /\Lambda_h)^2$. For $||v_a^{(0)}|| \gg 1$ the linear term in the potential is also important, which makes the potential steeper and shortens the duration of inflation.
For $||v_a|| \ll 1$ one finds additional inflation near the initial `critical' point, and simulations show that
the average number of e-folds is given by the sum of \eqref{eq:NeSmallv0} and \eqref{eq:NeLargev0}, up to a factor of at most three, for all $||v_a^{(0)}|| \leq1$.

\begin{figure}[ht!]
    \begin{center}
     \subfigure[]{
        \includegraphics[width=0.45\textwidth]{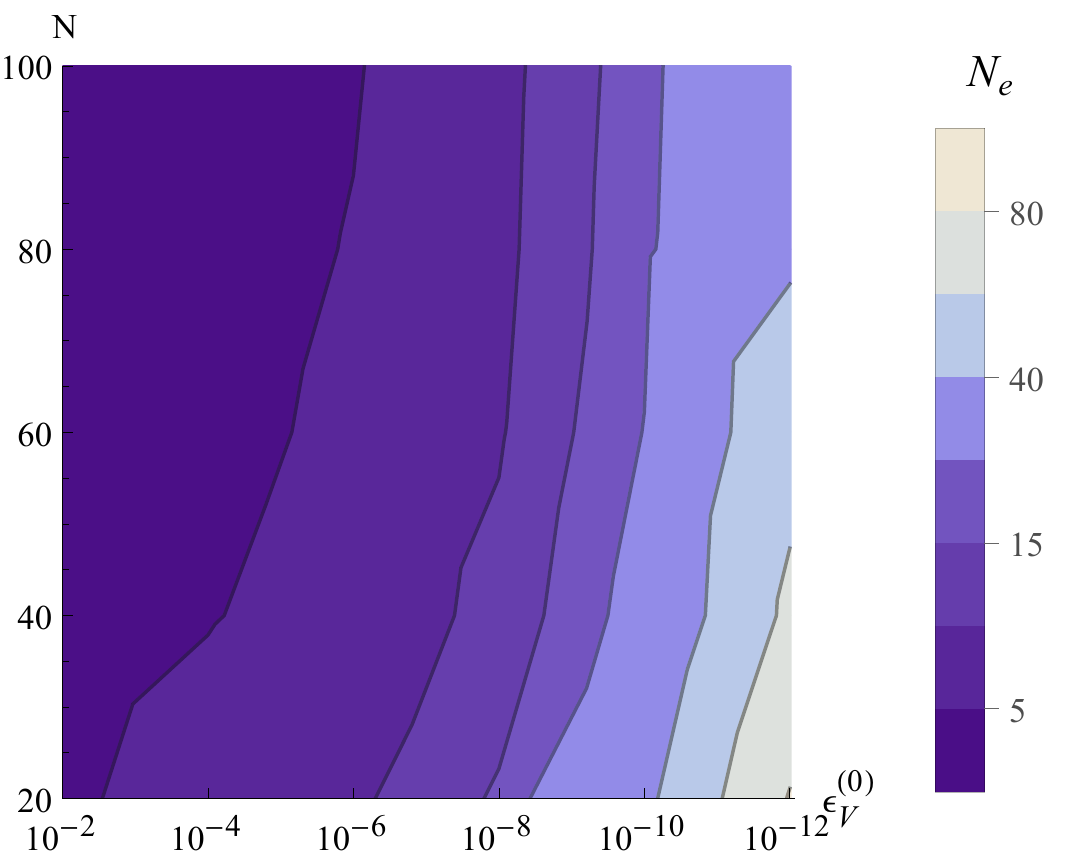}}
     \qquad
     \subfigure[]{
        \includegraphics[width=0.45\textwidth]{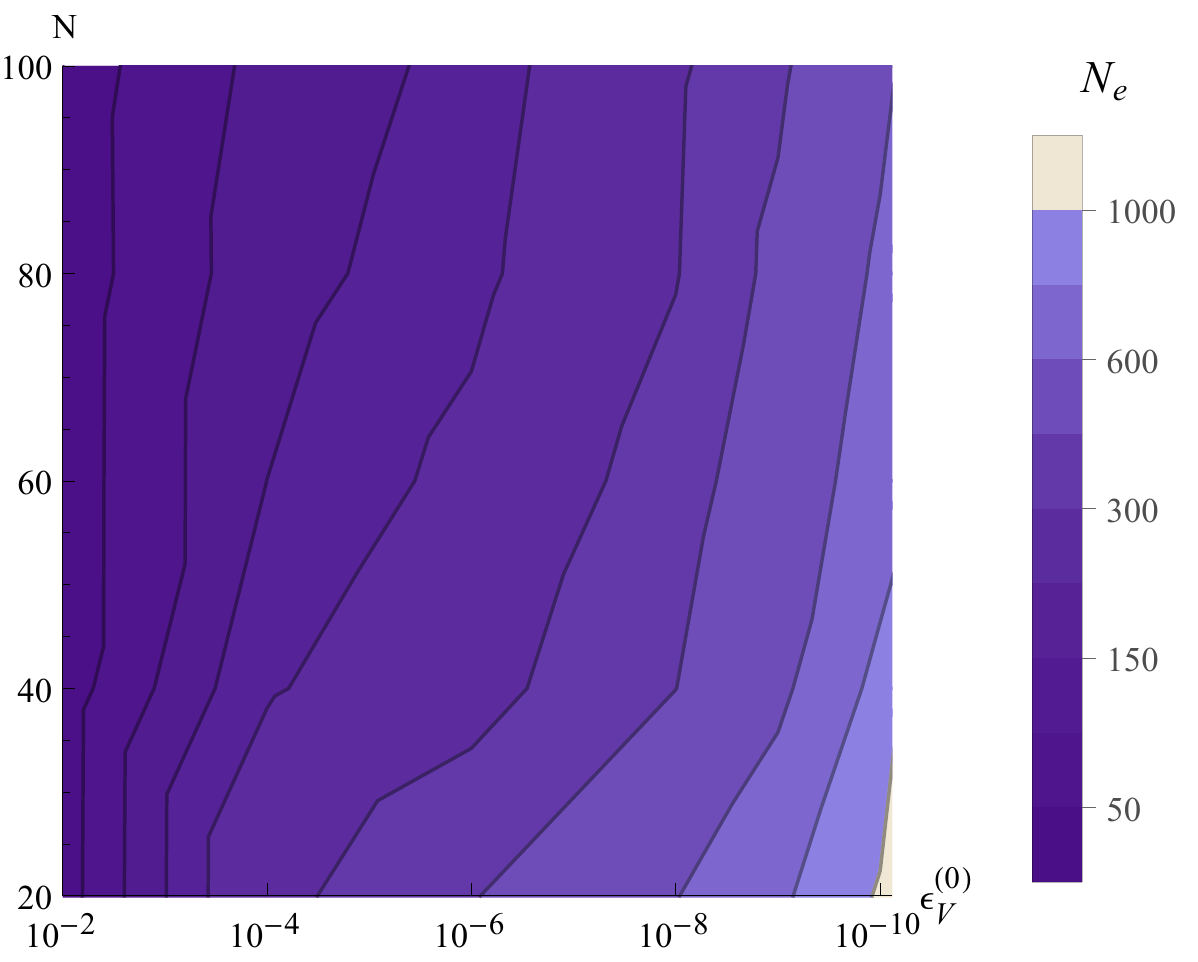}}
    \end{center}
    \caption{Contour plots of the number of e-folds $N_e$ for different values of $\epsilon_V^{(0)}$ and the matrix size $N$. Panel (a) is for $\frac{\Lambda_h}{M_p}= \frac{1}{100}$, $v_0 = \frac{10,000}{\sqrt{N}}$ and $\eta_V^{(0)}=10^{-6}$ while panel (b) is for $\frac{\Lambda_h}{M_p}= \frac{1}{10}$, $v_0 = \frac{10,000}{\sqrt{N}}$ and $\eta_V^{(0)}=10^{-6}$.}
   \label{fig:NvsEpsilon}
\end{figure}

In figure \ref{fig:NvsEpsilon} we plot the number of e-folds obtained in numerical simulations for different $N$, versus the initial value $\epsilon_V^{(0)}$. Figure \ref{fig:NvsEpsilon}(a) is for inflation near a critical point since $v_0 < (\M/\Lambda_h)^2$, and should be compared with our analytic formula \eqref{eq:NeSmallv0}. The parameters in \ref{fig:NvsEpsilon}(b) are chosen such that $v_0 \gg (\M/\Lambda_h)^2$. For $\epsilon_V^{(0)} \approx 10^{-4}$ we have $||v_a^{(0)}|| \approx 1$ and we can compare with \eqref{eq:NeLargev0}. For smaller values of $\epsilon_V^{(0)}$ we get a rough estimate for the number of e-folds by combining \eqref{eq:NeSmallv0} and \eqref{eq:NeLargev0}.

\subsection{Inflation from the Final Patch: $\Lambda_h \gtrsim \M$} \label{sec:LargeField}

For $\Lambda_h \gtrsim M_{Pl}$, the potential varies slowly over Planckian distances.\footnote{We note that previous constructions of \FRPs~have generally assumed $\Lambda_h \gg \M$.} The explicit construction of potentials in this class in string theory has proven to be challenging, and whether or not there exist well-controlled multifield theories of this form, with the potential varying slowly in every direction, is an open question. In this paper, we will not address the question of the ultraviolet completion of the theory, but rather briefly discuss the phenomenological properties of these potentials, and how they differ from potentials with $\Lambda_h\ll\M$.

To construct a large ensemble of potentials with $\Lambda_h \gtrsim \M$, the method of Dyson Brownian motion presented in this paper is still appropriate.  However, there is a regime in which a sizable fraction --- or all --- of the final 60 e-folds of inflation will occur in the final descent into a  minimum of the potential. However, in the implementation of Dyson Brownian motion presented in this work, a typical trajectory very rarely ends up in a minimum with a phenomenologically viable cosmological constant. Because of this, our approach is inefficient for studying these cases. On the other hand, during the final evolution a quadratic expansion of the potential around the minimum provides a sufficient approximation.

An important distinction is whether the approach to the minimum supports just a fraction of the last 60 e-folds, or all 60. In the first case, which occurs when $\Lambda_{h}\sim \M$, it is crucial to know the potential also away from the minimum, because it is there that cosmological scales have left the horizon. In the second case, occurring when $\Lambda_{h}\gg\M$, the characterization of the potential around the minimum is sufficient to determine the phenomenology. We can readily say a few things about this second case.

For $\Lambda_h \gg  \M$, we focus on the field dynamics as the field enters the patch surrounding the final minimum. The techniques of stationary random matrix theory reviewed in \S\ref{sec:rmt} are particularly well-suited to address this  problem, but since this would take us away from the main focus of this paper, we postpone a thorough investigation of the rich phenomenology ensuing from this ``single patch'' class of models. Below we briefly discuss just one interesting regime as an example of the dynamics in this class.

Expanded to quadratic order  around the final minimum, the potential  takes the form
\begin{equation}
V= V_0 + \frac{1}{2} m^2_{ab} \phi^a \phi^b = \Lambda_{v}^{4} \sqrt{N} \left[v_0  + \frac{1}{2}  v_{ab} \tilde \phi^a \tilde \phi^b  \right]\,,
\label{eq:LargeFieldPot}
\end{equation}
and to discuss a realistic inflationary phase that ends in a vacuum with a comparatively small cosmological constant, we will take $v_0$ to be negligible.
Because the kinetic terms are canonical, the mass matrix $v_{ab}$ can be diagonalized so that the dynamics in the final approach to the minimum is that of $N$ fields that are uncoupled in the scalar potential and only interact through the Hubble parameter. Coupling among the fields will be induced as one moves sufficiently far from the minimum, but in the following we will assume these effects are negligible. Notice that for large enough $\Lambda_{h}$ this can always be ensured.

The Hessian matrix at the minimum is an atypical member of the Gaussian Orthogonal Ensemble, with no negative eigenvalues.  As discussed in \S\ref{sec:fluct}, such a matrix can be regarded as a member of the GOE for which all eigenvalues have `fluctuated to positivity'.

The resulting inflationary evolution is that of {\it{assisted chaotic inflation}} \cite{Liddle:1998jc}. Each field slowly rolls down its own quadratic potential, feeling Hubble friction generated by all the other fields. The quantitative details of the dynamics depend on several factors.

As an example, the mass of the lightest field around the minimum is model-dependent.  If one allows for arbitrarily light fields (i.e.~requiring only positivity of the Hessian eigenvalues) it is possible that a very short distance $d$ away from the minimum (as compared with $\Lambda_{h}$) the eigenvalue relaxation discussed in \S\ref{sec:EigRelaxation} makes the lightest directions tachyonic. This in turn  would imply that the single-patch  quadratic approximation \eqref{eq:LargeFieldPot} breaks down a  distance $d \ll \Lambda_{h}$ from the minimum. This is a potentially interesting regime, but  a careful investigation of   classical and quantum mechanical effects driving the system away from the minimum would be necessary.
If we instead assume that there is a gap $\xi$ in the spectrum, then one can ensure that no tachyonic directions are generated for a substantial fraction of the distance $\Lambda_{h}$. More specifically, in this regime the spectrum around the minimum is given by (\ref{eq:fluct}) with $\xi>0$ (analogous to figure \ref{fig:WignerSpectra} but with a shifted left edge). For $\xi\sim \mathcal{O}(1)$ one finds that tachyonic directions develop only at a distance $\mathcal{O}(\Lambda_{h})$  from the minimum. In this case all $N$ fields have approximately the same masses, up to factors of order  unity. Inflation in a similar system has been studied in \cite{Dimopoulos:2005ac} and goes under the name of N-flation. When $\xi\ll 1$ then there is a considerable hierarchy in the masses of the fields. In this case, as inflation proceeds, the heavier fields settle down first and only the remaining lighter fields keep driving inflation, as discussed\footnote{Notice, however, that the analysis of \cite{EM05} considered the  displacement of the axionic fields from a  metastable KKLT  vacuum of type IIB  string theory,  for which $\Lambda_h < M_{Pl}$. Our discussion here is agnostic about a string theory realization, and crucially uses $\Lambda_h \gg M_{Pl}$.} e.g.~in \cite{EM05}.

To summarize, for $\Lambda_h \gtrsim \M$, a sizable fraction --- or all --- of the final 60 e-folds of inflation occur as the fields make their final descent into the minimum. Although qualitatively the relevant dynamics is that of assisted inflation, detailed predictions require careful consideration of the spectrum around the minimum, and of several other model-dependent choices. Random matrix theory can be used to characterize the properties of inflation in this regime, but we leave a detailed investigation for future work.

\section{Conclusions} \label{sec:concl}

In this paper, we have established a novel method for  constructing and analyzing random potentials for $N \gg 1$ scalar fields.  Our approach is to  describe the potential in the neighborhood of a path in field space by a series of quadratic approximations, with the Hessian matrix at each step determined by a stochastic process.

Demanding that the Hessian matrices at widely-separated points comprise a random sample of a rotationally-invariant statistical ensemble led us to Hessian matrices belonging to the Gaussian Orthogonal Ensemble.  The corresponding local, stochastic process for the Hessian is precisely Dyson Brownian motion.  Although we emphasized the simple case of a GOE landscape generated by Dyson Brownian motion in this work, our approach is far more general: taking the stochastic process for the Hessian to be a generalization of Dyson Brownian motion immediately leads to a broader class of random potentials.

We expect that the Dyson Brownian motion construction of a GOE landscape gives a reasonably accurate local model of potentials arising in random supergravity theories with $N \gg 1$ fields, in the regime where the supersymmetry-breaking masses are larger than the supersymmetric masses.  In the complementary regime where the supersymmetric masses are important, significant correlations in the Hessian matrix make the eigenvalue statistics differ from that of the GOE \cite{Denef:2004cf,Marsh:2011aa}.  This regime is also accessible by the methods described in this work, but the stochastic process will differ from Dyson Brownian motion.  Using stochastic evolution of the Hessian to study potentials with spontaneously broken supersymmetry is an interesting task for the future.

Our method for constructing a random potential is  clearly useful for answering questions that depend on local information in the vicinity of a path, but its utility for genuinely global problems, such as counting minima, is  far from obvious.   It would be worthwhile to  investigate  extensions that make use of  local, stochastic evolution of the Hessian, yet  extract enough global information to address more general  questions.

An important application  of our method is the characterization of inflation
in theories with $N \gg 1$ coupled scalar fields.  For the ensemble of potentials generated by Dyson Brownian motion, non-equilibrium random matrix theory provides powerful new tools for numerical and analytical studies of these systems.
We have
examined
the inflationary dynamics ---
at the level of the homogeneous background --- up to systems of size $N=100$, which is an order of magnitude larger than achieved with prior constructions. Extending our work to even larger systems  should be straightforward.

The nature of inflation in an ensemble of  Dyson Brownian motion potentials depends on the correlation length $\Lambda_h$.
We analyzed a large class of multifield Dyson Brownian motion potentials with structure on sub-Planckian scales, which supported a period of inflation as a consequence of an initial fine-tuning of the local gradient and mass spectrum.
In this regime, eigenvalue repulsion induces a rapid relaxation of the spectrum of the Hessian matrix, rapidly undoing any initial fine-tuning of the $\eta$ parameter.  As a result,  prolonged inflation is very rare,  even if the slow roll parameters are small at the beginning of the trajectory.\footnote{The likelihood that a given critical point will have small enough curvature to support the {\it{onset}} of inflation has been determined, for GOE statistics, in \cite{Pedro:2013nda}.}

For potentials varying slowly over Planckian distances (corresponding to $\Lambda_h \gtrsim M_{Pl}$), which could presumably arise if each scalar were protected by an approximate shift symmetry respected by the ultraviolet theory up to the scale $\Lambda_h$, some of the final 60 e-folds  are typically generated in the final approach to a minimum.   A  modified implementation of Dyson Brownian motion could be used to study inflation in this regime, but is beyond the scope of this work.

Our explicit construction of an ensemble of potentials with
many coupled scalar fields offers an unprecedented opportunity to
compute the primordial perturbations in inflationary models with $N \gg 1$  fields.
As the prevalence of instabilities makes it plausible that entropic perturbations will grow toward the end of inflation,  the approach to an adiabatic limit (cf.~\cite{Meyers:2010rg,Seery:2012vj}) may be quite complicated, and
determining the predictions of
multifield inflation in general potentials remains an important problem for the future.

\section*{Acknowledgements}

We thank Thomas Bachlechner, Daniel Baumann, Mafalda Dias, Richard Easther, Jonathan Frazer, S\'ebastien Renaux-Petel, Henry Tye, Dan Wohns, Gang Xu,  and Itamar Yaakov for discussions related to this work.  We are grateful to Francisco Pedro and Alexander Westphal for useful discussions of \cite{Pedro:2013nda}.  The research of L.~M.~was  supported in part by the NSF under  grant PHY-0757868. E.~P.~was supported in part by the Department of Energy grant DE-FG02-91ER-40671. T.~W.~was supported by a Research Fellowship (Grant number WR 166/1-1) of the German Research Foundation (DFG).

\bibliographystyle{modifiedJHEP}
\bibliography{refs}

\end{document}